\documentclass[twocolumn, superscriptaddress, nobibnotes, nofootinbib, aps, pra]{revtex4-2}
\usepackage{graphicx}
\usepackage{amssymb}
\usepackage{amsmath}
\usepackage{subcaption}
\usepackage{braket}
\usepackage{multirow}
\usepackage{tikz}
\usetikzlibrary{quotes, angles}
\usepackage{upgreek}
\usepackage{hyperref}
\hypersetup{colorlinks=true,linkcolor=blue,filecolor=blue,citecolor = blue, urlcolor=cyan}
\usepackage{mathbbol}

\usepackage{lipsum}

\begin{document}

\title{Destructive Interference and Inertial Noise in Matter-wave Interferometry}

\author{Meng-Zhi Wu}
    \email{mengzhi.wu@rug.nl}
    \affiliation{Van Swinderen Institute for Particle Physics and Gravity, University of Groningen, 9747 AG Groningen, the Netherlands }

\author{Marko Toro\v{s}}
%    \affiliation{School of Physics and Astronomy, University of Glasgow, Glasgow, G12 8QQ, UK}
%    \affiliation{Department of Physics and Astronomy, University College London, Gower Street, WC1E 6BT London, UK}
    \affiliation{Faculty of Mathematics and Physics, University of Ljubljana, Jadranska 19, SI-1000 Ljubljana, Slovenia}

\author{Sougato Bose}
    \affiliation{Department of Physics and Astronomy, University College London, Gower Street, WC1E 6BT London, UK}

\author{Anupam Mazumdar} 
    \email{anupam.mazumdar@rug.nl}
    \affiliation{Van Swinderen Institute for Particle Physics and Gravity, University of Groningen, 9747 AG Groningen, the Netherlands }

\begin{abstract}
    Matter-wave interferometry is highly susceptible to inertial acceleration noises arising from the vibration of the experimental apparatus. There are various methods for noise suppression. In this paper, we propose leveraging the cross-correlation of multi-directional vibration noises to mitigate their dephasing effect in matter-wave interferometers. Specifically, we analyse an interferometer driven by its internal state under an external field and examine the dephasing caused by a two-dimensional random inertial force. As we will demonstrate, the coupling between the two-dimensional inertial force noise components will shift the resonance peak but not change the shape of the power spectral density. Moreover, when the noise approximately resonates with the intrinsic frequency of the test mass, we find that the standard deviation of the phase can be suppressed by a factor roughly equal to the Q-factor of the noise. This technique holds significant potential for future gravity experiments utilising quantum sensors, such as measuring gravitational acceleration and exploring quantum entanglement induced by gravity. 
\end{abstract}

\maketitle

\section{Introduction}\label{section 1}

Matter-wave interferometry has numerous prominent applications for gravity experiments, such as measuring the gravitational acceleration and the gravity gradient~\cite{Peters1999, science.1135459, Stray2022}, testing the equivalence principle~\cite{PhysRevLett.120.183604, PhysRevLett.125.191101, Bose:2022czr}, detecting gravitational waves~\cite{refId0, Marshman:2018upe,Junca_2019, Abe_2021, Mitchell:2022zbp} and exploring the quantum nature of gravity~\cite{PhysRevLett.119.240401,ICTS, PhysRevLett.119.240402,Biswas:2022qto,Hanif:2023fto}.

Robustness is a critical challenge for current and future interferometers as various quantum fluctuations can cause the decoherence of the test mass~\cite{Joos_Zeh_1985,Zeh:1970zz,schlosshauer2007quantum,Schut:2024lgp,Hornberger_2012}. The arguably simplest example is given by the decoherence of a qubit system, which can be described by the Bloch-Redfield equation containing both dephasing and relaxation~\cite{10.1063/1.5089550, doi:10.1080/00018730802218067}.  Moreover, even classical noises in the modelling can induce decoherence of the internal qubit space as well as a loss of interferometric contrast~\cite{LucBouten_2004}. 

Regarding the spatial degrees of freedom, there are several distinct manifestations of decoherence. Firstly, spatial decoherence can lead to a dephasing effect, characterized by a decay factor $\mathrm{e}^{-\Gamma t}$ arising from the ensemble average of the random phase factor $\mathbb{E}[\mathrm{e}^{i\delta\phi}]$, which can inhibit the readout of the internal qubits~\cite{Wu:2024tcr,bowen2015quantum}, here $\Gamma$ can be treated as a constant. Secondly, the ensemble average of the noise can also lead to a spatial dissipator $\Lambda\left[\hat{x}, \left[\hat{x}, \hat{\rho}\right]\right]$ in the master equation, which contributes a decoherence factor $\mathrm{e}^{-\Lambda(x_1-x_2)^2}$ on the density matrix in position space~\cite{PhysRevA.84.052121, RomeroIsart2011LargeQS, RieraCampeny2024wigneranalysisof}. In addition, due to the fluctuation of the trajectories of the superpositions, noises can also lead to a non-closure problem of the test mass, which can also lead to loss of qubit witness when tracing out the spatial degrees of freedom, known as the Humpty-Dumpty problem~\cite{Englert1988, Schwinger1988, PhysRevA.40.1775}. 

Physically there are many sources of noises such as vibration, inertial forces~\cite{PhysRevD.111.064004}, Coulomb/dipole interactions~\cite{PhysRevA.110.022412,Fragolino:2023agd}, current/magnetic field fluctuations~\cite{Moorthy:2025fnu}, and gravity fluctuations~\cite{PhysRevResearch.3.023178,Schut:2023eux}, which couple to the test mass in different ways.

For acceleration noises which couple to the system linearly, the position localisation decoherence and the Humpty-Dumpty problems vanish due to common mode noise cancellation, see~\cite{Wu:2024tcr}. On the other hand, the phase fluctuation caused by this type of noise is precisely the path integral of the Lagrangian of the noise along the undisturbed trajectories of the interferometer~\cite{Storey1994TheFP}. Consequently, the dephasing factor can be regarded as a linear response to the noise, and the transfer function is the Fourier transform of the ideal differential trajectory~\cite{Wu:2024tcr, PhysRevD.111.064004, PhysRevD.107.104053,Moorthy:2025fnu}. 

To suppress the noise-induced dephasing effect, a commonly used method is to use the cross correlation between the noise and a series of control pulses. An alternative innovative strategy is to use the destructive interference among several correlated noises~\cite{PhysRevLett.132.223601, Cai_2012, Stark2017, Cohen2016}. For uncorrelated noises, it is principally possible to introduce a coupling among them to mitigate the consequent dephasing by destructive interference. 

Therefore, we will aim to theoretically investigate this possibility for multi-dimensional acceleration noises in matter-wave interferometers.
We will consider a test mass under the influence of two-dimensional inertial forces, physically arising from the vibration of the experimental apparatus. 
The correlation between the two-dimensional noise components, corresponding to the vibrations along different directions of the experimental apparatus, will be introduced phenomenologically. This can be experimentally achieved using a vibration direction converter~\cite{Ito1972StudyOR, xu2018equivalent}.
Consequently, this correlation will induce destructive interference between the noises along different directions, resulting in the suppression of the overall dephasing of the interferometer.

This paper is organised as follows. In section 2, we will describe the ideal dynamics of a two-dimensional interferometer driven by its qubit without noise. In section 3, we will construct the general theory of the dephasing effect due to two-dimensional noise and provide a comprehensive analysis of the noise suppression. In section 4, we will study a typical source of noise as a phenomenological example of the general theory, which is the inertial acceleration noise modelled by the two-dimensional Langevin equation. In section 5, we will continue to discuss the dephasing factor and noise suppression of the two-dimensional inertial noise. In section 6, we will discuss the application of this proposal to gravity experiments.

\section{interferometeric setup}\label{section 2}

In this paper, we consider a matter-wave interferometer that is initially trapped in a quadratic potential and driven by a state-dependent force, which is described by the following Hamiltonian~\footnote{In principle, there can be some coupling terms such as the rotation in the $x$-$y$ plane, described by the Hamiltonian
\begin{equation*}
    \hat{H}_{\rm rot} = \omega_{\rm rot}(\hat{x}\hat{p}_y - \hat{y}\hat{p}_x),
\end{equation*}
which is exactly the $z$-component orbital angular momentum $\hat{L}_z$. This term has a significant effect on the nanoparticle while describing the one-loop interferometry, see \cite{PhysRevLett.130.113602, Zhou:2024pdl, Rizaldy:2024viw}.}
\begin{equation}\label{EOM}
    \hat{H}_{\rm tot} = \frac{\hat{p}_x^2+\hat{p}_y^2}{2m} + \frac{1}{2}m\omega_0^2(\hat{x}^2+\hat{y}^2) + g_c(\hat{\sigma}_x\hat{x} + \hat{\sigma}_y\hat{y}) + \hat{H}_{\rm qubit}, 
\end{equation}
where $m$ is the test mass used in the interferometer, $\omega_0$ is the intrinsic frequency of the trap, $g_c(\hat{\sigma}_x\hat{x}+\hat{\sigma}_y\hat{y})$ is the coupling between the internal state and the spatial degree of freedom~\footnote{The coupling factor $g_c$ physically originates from various mechanisms. For example, the $g_c$ for a Stern-Gerlach interferometer is the magnetic field gradient, i.e. $g_c\sim\partial B/\partial x$.}, and $\hat{H}_{\rm qubit}$ is the Hamiltonian of the internal state, which sometimes can be also referred to as the qubit of the test mass. This state-driven interferometer can be implemented by "artificial" spin-orbit coupling~\cite{PhysRevA.109.063310, Yoneda2018}, SQUID~\cite{Johnsson2016}, Stern-Gerlach scheme~\cite{PhysRevLett.111.180403,PhysRevLett.125.023602,PhysRevResearch.4.023087}. 
The internal state Hamiltonian $\hat{H}_{\rm qubit}$ can be usually formed as 
\begin{equation}
    \hat{H}_{\rm qubit} = \hbar\omega_q\sum\limits_{i=x,y,z}n_i\hat{\sigma}_i = \hbar\omega_q\hat{\sigma}_n,
\end{equation}
at the leading order, where $(n_x, n_y, n_z)$ defines a certain $n$-direction in the Bloch sphere of the internal state. In this paper, we assume that $\hat{H}_{\rm qubit}$ is much larger than the coupling term $g_c\hat{\mathbf{\sigma}}\cdot\hat{\mathbf{r}}$ so that the motion of the internal space is dominant by $\hat{H}_{\rm qubit}$. Consequently, the internal states are fixed as the eigenstates of $\hat{S}_n$ without transition during the experiment, which will be denoted as $\ket{\uparrow}$ and $\ket{\downarrow}$, i.e. $\hat{\sigma}_n\ket{\uparrow}=\ket{\uparrow}$ and $\hat{\sigma}_n\ket{\downarrow}=-\ket{\downarrow}$.

As a result, we can use the expectation values $\braket{\uparrow|\hat{\sigma}_i|\uparrow}$ and $\braket{\downarrow|\hat{\sigma}_i|\downarrow}$ of the internal state to replace the Pauli matrices in the original Hamiltonian. We further assume the $n$-axis is confined in the $x$-$y$ plane for ease of analysis, i.e.
\begin{equation}\label{angle theta}
    (n_x, n_y, n_z)=(\cos\theta, \sin\theta, 0),
\end{equation}
with a deflection angle $\theta$, then the expectation values are
\begin{equation}
\begin{aligned}
    \braket{\uparrow|\hat{\sigma}_x|\uparrow} = \cos\theta, \quad \braket{\downarrow|\hat{\sigma}_x|\downarrow} = -\cos\theta \\    
    \braket{\uparrow|\hat{\sigma}_y|\uparrow} = \sin\theta, \quad \braket{\downarrow|\hat{\sigma}_y|\downarrow} = -\sin\theta.
\end{aligned}
\end{equation} 
Thus, we can focus on the spatial degrees of freedom of the test mass, and the spatial Hamiltonian of the test mass becomes
\begin{equation}
    \hat{H} = \frac{\hat{p}_x^2+\hat{p}_y^2}{2m} + \frac{1}{2}m\omega_0^2(\hat{x}^2+\hat{y}^2) + m(a_x\hat{x}+a_y\hat{y}),
\end{equation}
where the driving forces arise from
\begin{eqnarray}
    ma_x=&\pm g_c\cos\theta\,,~~~
    ma_y=&\pm g_c\sin\theta
\end{eqnarray}
are determined by the eigenvalues $\pm 1$ of the Pauli matrices. Then the time evolution operator for each path has a form~\cite{Wu:2024tcr} 
\begin{equation}
    \hat{U}_\pm(t) = \mathrm{e}^{i\phi_\pm(t)}\hat{D}(\alpha_\pm(t))\exp\left[-\frac{i}{\hbar}\hat{H}_0t\right]\hat{D}(-\alpha_\pm(0)),
\end{equation}
where $\hat{H}_0$ is the Hamiltonian of a 2-dimensional simple harmonic oscillator without driven forces, $\phi_\pm(t)$ is the path integral phase along the classical trajectories, given by
\begin{equation}
    \phi_\pm(t) = \frac{m}{\hbar}\int_0^t \frac{1}{2}(\dot{x}_\pm^2+\dot{y}_\pm^2) - \frac{1}{2}\omega_0^2(x_\pm^2+y_\pm^2) - (a_xx_\pm+a_yy_\pm)\,\mathrm{d}t,    
\end{equation}
and $\hat{D}(\alpha_\pm(t))$ is the displacement operator with $\alpha_\pm(t)$ as the trajectories in the phase space, i.e.
\begin{equation}
    \hat{D}(\alpha(t)) = \exp\left[\alpha_x(t)\hat{b}_x^\dagger-\alpha_x^*(t)\hat{b}_x\right] \exp\left[\alpha_y(t)\hat{b}_y^\dagger-\alpha_y^*(t)\hat{b}_y\right],
\end{equation}
where $\hat{b}_x$ and $\hat{b}_y$ are the annihilation operators of $\hat{x}$ and $\hat{y}$, and 
\begin{equation}
    \alpha_x(t)=\sqrt{\frac{m\omega_0}{2\hbar}}\left(x(t)+ip_x(t)/m\omega_0\right),
\end{equation}
is the trajectory in the classical phase space. Notably, displacement operators don't change the shape of the wavefunction of the test mass (for a test mass initially in a mixed state, the shape of its Wigner function also doesn't change), so the time-evolution of the system can be determined by the classical trajectories~\cite{Wu:2024tcr}.

\begin{figure}
    \centering
    \includegraphics[scale=0.5]{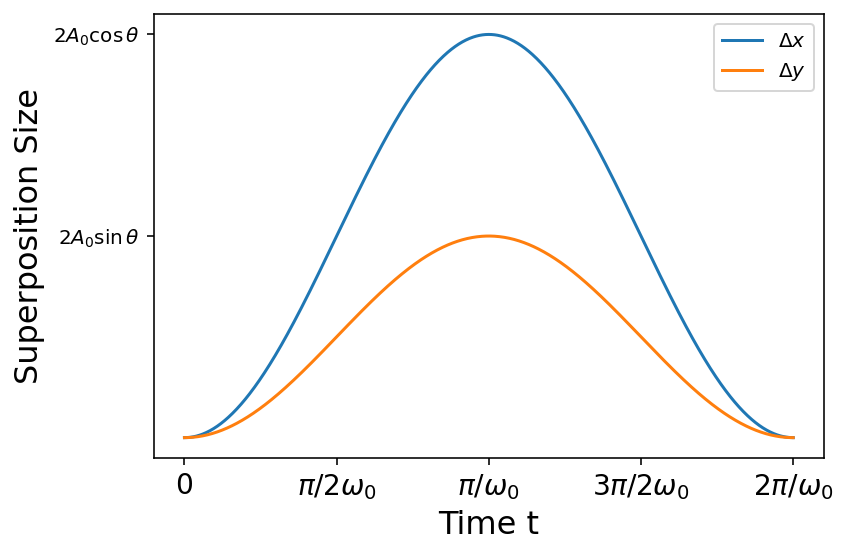}
    \caption{\small The superposition size of a 2-dimensional interferometer driven by its internal state. Both the $x$- and $y$-directions follow trigonometric functions $\sim(1-\cos\omega_0t)$ with different amplitudes, i.e. $\Delta x\sim2A_0\cos\theta$ and $\Delta y\sim2A_0\sin\theta$. The period of the interferometer is $2\pi/\omega_0$, where the angular frequency $\omega_0$ can be set to various values for different experiments. }
    \label{trajectory}
\end{figure}

For simplicity, we consider constant driven forces, and then each path of the interferometer is a simple harmonic oscillation with a fixed displacement $(a_x/\omega_0^2, a_y/\omega_0^2)$. In this case, the coordinates in phase space are simply 
\begin{equation}
    \alpha_{x\pm}(t)=(\mathrm{e}^{-i\omega_0t}-1)a_x\,,~~\alpha_{y\pm}(t)=(\mathrm{e}^{-i\omega_0t}-1)a_y.
\end{equation}
Consequently, the time evolution of a test mass initially prepared in the ground state is given by
\begin{equation} 
    \ket{\psi(t)} = \frac{\mathrm{e}^{i\phi_+(t)}\ket{\alpha_{x+}(t), \alpha_{y+}(t)} + \mathrm{e}^{i\phi_-(t)}\ket{\alpha_{x-}(t), \alpha_{y-}(t)}}{\sqrt{2}}.
\end{equation}
Then the 2-dimensional classical trajectories of the interferometer are simply harmonic oscillations and given by:
\begin{equation}\label{ideal trajectories}
\left\{\begin{aligned}
    x_{\pm}(t) &= \pm A_0\cos\theta(1-\cos\omega_0t), \\
    y_{\pm}(t) &= \pm A_0\sin\theta(1-\cos\omega_0t),
\end{aligned}\right.
\end{equation}
where $A_0=g_c/(m\omega_0^2)$ is the oscillation amplitude due to the driven force. Thus, the differential trajectories are
\begin{equation}
\left\{\begin{aligned}
    x_+(t)-x_-(t) &= 2A_0\cos\theta(1-\cos\omega_0t), \\
    y_+(t)-y_-(t) &= 2A_0\sin\theta(1-\cos\omega_0t).
\end{aligned}\right.
\end{equation}
The 2-dimensional superposition size is shown in Fig.~\ref{trajectory}. The values of parameters remain symbolic, as they vary across different experiment setups. For instance, $\omega_0$ is set to $\sim100$\,kHz in \cite{PhysRevLett.111.180403} and $\sim1$\,Hz in \cite{PhysRevResearch.4.023087}.

In an experiment, the observable quantity is usually constructed by the differential phase of the two arms, i.e. 
\begin{equation}
    \phi_{\rm diff} \equiv \phi_+-\phi_-,    
\end{equation}
which is usually measured by Ramsey interferometry~\cite{PhysRevLett.111.180403}. In the remaining sections, we will first examine the fluctuation of this differential phase under a generic noise and then focus on strategies for mitigating inertial noise.

\section{Dephasing of Generic Noise}\label{section 3}

In a real experiment, the test mass is always affected by external noises, which can cause decoherence of the test mass via several mechanisms. The most typical noise is the acceleration noise, which couples the system linearly, such as the vibration, the Coulomb interaction~\cite{PhysRevA.110.022412} and the inertial force~\cite{PhysRevResearch.3.023178}. This type of noise only causes a dephasing effect, while the position localisation decoherence and contrast loss effect can be cancelled, because this type of noise affects both arms in the same way~\cite{PhysRevResearch.3.023178, Wu:2024tcr}.

The Hamiltonian for the 2-dimensional acceleration noise can be generally written as
\begin{equation}
    \hat{H}_{\rm noise} = m\left(\delta a_x(t)\hat{x}+\delta a_y(t)\hat{y}\right),
\end{equation}
where $\delta a_x(t)$ and $\delta a_y(t)$ are the acceleration noises along $x$- and $y$-direction, which are usually formulated as two stationary Gaussian processes with zero mean values. In addition, the Wiener-Khinchin theorem~\cite{Khintchine1934,Wiener:1930} states that their auto time-correlation equals the Fourier transform of their \emph{power spectral density} (PSD) 
\begin{equation}
    S_{a_ia_j}(\omega)\equiv\lim_{T\to\infty}\frac{\mathbb{E}[\delta\tilde{a}_i(\omega)\delta\tilde{a}_j^*(\omega)]}{T},
\end{equation}
where $\delta\tilde{a}_i(\omega)$ is the Fourier transform of $\delta a_i(t)$ over the finite time domain $0\sim T$ and $\mathbb{E}[\cdot]$ denotes the ensemble average of a stochastic variable. In sum, the expectation value and the time correlation of $\delta a_i(t)$ satisfy
\begin{equation}
\begin{aligned}
    \mathbb{E}[\delta a_i(t)] &= 0,\quad \forall t, \\
    \mathbb{E}[\delta a_i(t_1)\delta a_j(t_2)] &= \int S_{a_ia_j}(\omega)\mathrm{e}^{-i\omega(t_2-t_1)}\,\mathrm{d}\omega,\quad \forall t_1,~t_2.
\end{aligned}
\end{equation}
The fluctuation of the differential phase due to this external noise can be formulated as the path integral of the noise term along the unperturbed trajectories~\cite{Wu:2024tcr, Storey1994TheFP}, that is
\begin{equation}\label{delta phi}
    \delta\phi_{\rm diff} = \frac{1}{\hbar}\int m\left(\delta a_x(x_R-x_L)+\delta a_y(y_R-y_L)\right)\,\mathrm{d}t.
\end{equation}
Since the noises are assumed to be Gaussian, the phase fluctuation $\delta\phi_{\rm diff}$ also follows a Gaussian distribution with a zero mean value. Moreover, the ensemble average of this random phase factor contributes a decay factor characterized by its variance $\sigma_{\phi_{\rm diff}}^2 \equiv \mathbb{E}[(\delta\phi_{\rm diff})^2]$~\footnote{This result can be obtained by directly computing the probability integral 
\begin{equation*}
    \mathbb{E}[\mathrm{e}^{i\delta\phi_{\rm diff}}] = \int \mathrm{e}^{i\delta\phi_{\rm diff}}\frac{1}{\sqrt{2\pi}\sigma_{\phi_{\rm diff}}}\mathrm{e}^{-\frac{(\delta\phi_{\rm diff})^2}{2\sigma_{\phi_{\rm diff}}^2}}\,\mathrm{d}(\delta\phi_{\rm diff}).
\end{equation*}}, i.e.
\begin{equation}
    \mathbb{E}[\mathrm{e}^{i\delta\phi_{\rm diff}}] = \mathrm{e}^{-{\sigma_{\phi_{\rm diff}}^2}/{2}}.
\end{equation}
This decay factor can cause a purity loss $\Delta\mathcal{P}\approx\sigma_{\phi_{\rm diff}}^2/2$ and an entropy increase $\Delta S\approx \sigma_{\phi_{\rm diff}}^2/2$ of the internal state~\cite{Wu:2024tcr}. From the perspective of information theory, the information of the internal state gets lost due to the randomness of the noise, even if the noise is fundamentally classical. Due to the decoherence of the internal state, the witness of the Ramsey interferometry also gets lost, characterized by the factor $\mathrm{e}^{-\sigma_{\phi_{\rm diff}}^2/2}$~\cite{Wu:2024tcr}.

The variance can be directly written as the following integral
\begin{eqnarray}
    &\sigma_{\phi_{\rm diff}}^2 = \mathbb{E}[(\delta\phi_{\rm diff})^2]=\nonumber \\
    &\iint\sum\limits_{r_i,r_j=x,y}\mathbb{E}[\delta a_i(t_1)\delta a_j(t_2)]\Delta r_i(t_1)\Delta r_j(t_2)\,\mathrm{d}t_1\mathrm{d}t_2,
\end{eqnarray}
where $\Delta r_i(t)\equiv r_{iR}(t)-r_{iL}(t)$ with $r_i=x,y$. 

According to the Wiener-Khinchin theorem \cite{Wiener:1930,Khintchine1934}, $\mathbb{E}[\delta a_i(t_1)\delta a_j(t_2)]$ is the Fourier transform of $S_{a_ia_j}(\omega)$, then one may obtain
\begin{equation}\label{generic sigma2}
\begin{split}
    \sigma_{\phi_{\rm diff}}^2 &= \frac{m^2}{\hbar^2}\int S_{a_xa_x}(\omega)F_{xx}(\omega) + S_{a_ya_y}(\omega)F_{yy}(\omega) \\ 
        &+ S_{a_xa_y}(\omega)F_{xy}(\omega) + S_{a_ya_x}(\omega)F_{yx}(\omega) \,\mathrm{d}\omega.
\end{split}
\end{equation}
So the phase fluctuation can be regarded as a linear response to the noise, and the \emph{transfer functions} $F_{ij}(\omega)$ are defined as:
\begin{equation}\label{transfer function definition}
\begin{split}
    F_{ij}(\omega) &= \iint\Delta r_i(t_1)\Delta r_j(t_2)\mathrm{e}^{i\omega(t_1-t_2)}\,\mathrm{d}t_1\mathrm{d}t_2 \\
        &= \Delta\tilde{r}_i(\omega)\Delta\tilde{r}_j^*(\omega),
\end{split}
\end{equation}
where $\Delta\tilde{r}_i(\omega)$ is the Fourier transform of $\Delta r_i(t)$. It is remarkable that the transfer functions only depend on the undisturbed trajectories of the interferometer and they are independent of the external noises.

\begin{figure}
    \centering
    \includegraphics[scale=0.5]{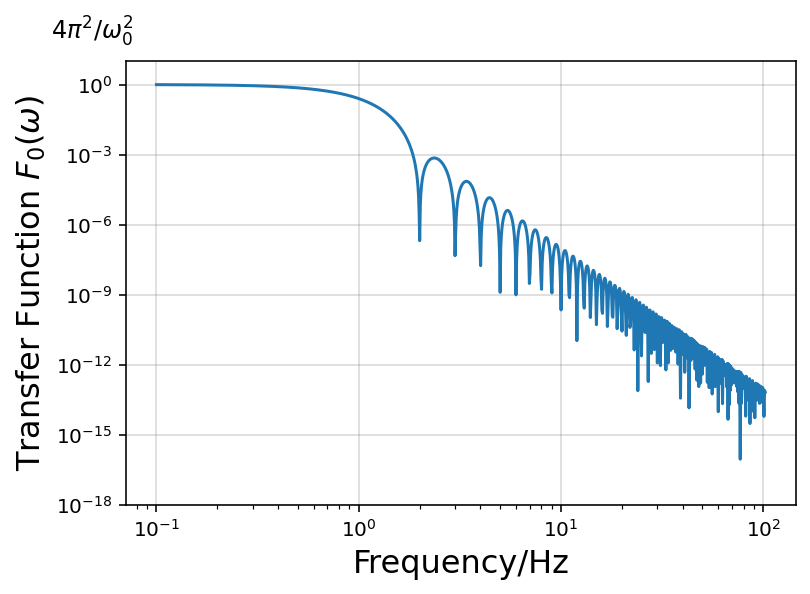}
    \caption{\small The dimensionless transfer function $F_0(\omega)$. It approaches a constant $4\pi^2/\omega_0^2$ in the low-frequency limit $\omega\to0$ and decreases at a rate proportional to $\omega^{-6}$ in the high-frequency limit $\omega\to\infty$.}
    \label{transfer}
\end{figure}

For the trajectories, Eq.\eqref{ideal trajectories}, the transfer functions can be simplified as
\begin{equation}
\begin{aligned}
    F_{xx}(\omega) &= 4A_0^2\cos^2\theta F_0(\omega), \\
    F_{yy}(\omega) &= 4A_0^2\sin^2\theta F_0(\omega), \\
    F_{xy}(\omega) &= F_{yx}(\omega) = 4A_0^2\sin\theta\cos\theta F_0(\omega),
\end{aligned} 
\end{equation}
where the dimensionless transfer function $F_0(\omega)$ is given by 
\begin{equation}\label{F0}
    F_0(\omega) \equiv \left|\int_0^{\frac{2\pi}{\omega_0}}(1-\cos\omega_0t)\mathrm{e}^{i\omega t}\,\mathrm{d}t\right|^2 = \frac{4\omega_0^4\sin^2\frac{\pi\omega}{\omega_0}}{\omega^2(\omega^2-\omega_0^2)^2}.
\end{equation}
As is shown in Fig. \ref{transfer}, $F_0(\omega)$ tends to a constant $4\pi^2/\omega_0^2$ in the low-frequency limit $\omega\to0$ and decreases with respect to $\omega^{-6}$ in the high-frequency limit $\omega\to\infty$.
Note that the cross PSD $S_{a_ya_x}(\omega)$ is the complex conjugate of $S_{a_xa_y}(\omega)$, so only the real part of $\bar{S}_{a_xa_y}(\omega)\equiv(S_{a_xa_y}(\omega)+S_{a_ya_x}(\omega))/2$, known as the \emph{co-spectrum} or the \emph{in-phase} component of $S_{a_xa_y}(\omega)$, affects the variance of the phase fluctuation. By contrast, $\sigma_{\phi_{\rm diff}}^2$ is independent with the imaginary part of the cross PSD, often referred to as the \emph{quadrature spectrum} or the \emph{out-of-phase} component of $S_{a_xa_y}(\omega)$, which indicates the correlation between $\delta a_x$ and $\delta a_y$ with $90^{\circ}$-phase shift. Thus, $\sigma_{\phi_{\rm diff}}^2$ can be simplified: 
\begin{equation}
\begin{split}
    \sigma_{\phi_{\rm diff}}^2 &= \frac{4m^2}{\hbar^2}A_0^2\int \bigg[\cos^2\theta S_{a_xa_x}(\omega) + \sin^2\theta S_{a_ya_y}(\omega) \\ 
        &+ \sin2\theta\bar{S}_{a_xa_y}(\omega)\bigg] F_0(\omega)\,\mathrm{d}\omega.
\end{split}
\end{equation}
Furthermore, if the noise along $x$- and $y$-directions are isotropic, i.e. $S_{a_xa_x}(\omega)\equiv S_{a_ya_y}(\omega)\equiv S_{aa}(\omega)$, then the variance can be further simplified:
\begin{equation}\label{sigma integral}
    \sigma_{\phi_{\rm diff}}^2 = \frac{4m^2}{\hbar^2}A_0^2 \int\left[S_{aa}(\omega) + \sin2\theta\bar{S}_{a_xa_y}(\omega)\right]F_0(\omega)\,\mathrm{d}\omega.
\end{equation}
This integral can be computed by the residue theorem. As shown in Appendix \ref{appendix c}, the variance is given by:
\begin{equation}\label{sigma simplified}
\begin{split}
    \sigma_{\phi_{\rm diff}}^2 =& \frac{8\pi^2m^2}{\hbar^2\omega_0}A_0^2 \big[ S_{aa}(\omega_0) + \sin2\theta\bar{S}_{a_xa_y}(\omega_0) \\
        &+ 2S_{aa}(\omega\to0) + 2\sin2\theta\bar{S}_{a_xa_y}(\omega\to0) \big].
\end{split}
\end{equation}
Physically, it is interpreted that the test mass only resonates with the noise at the frequency $\omega_0$ and the zero-frequency, and it is orthogonal to the noise at other frequencies. It is because the trajectories $\Delta x(t)$ and $\Delta y(t)$ only have one frequency component $\omega_0$ and a constant bias.

For the zero-frequency resonance $S_{aa}(\omega\to0)$ and $S_{a_xa_y}(\omega\to0)$, we make several remarks as follows:
\begin{itemize}
    \item In a real experiment, the PSD and cross-PSD at zero frequency are not measurable because of the finite time domain which presents a natural cutoff $\omega_{\rm min}=2\pi/T_{\rm tot}$ in the low-frequency limit. Thus, the limit $\omega\to0$ can be exactly represented by this cutoff, $\omega_{\rm min}$.
    \item This resonance arises from the initial condition and the constant driving force. If a time-varying external force drives the superposition at a frequency $\omega_{\rm drive}$, then the test mass will resonate with $S_{aa}(\omega_{\rm drive})$ rather than $S_{aa}(\omega\to0)$.
\end{itemize} 
According to Eq.\eqref{sigma integral}, $\sigma_{\phi_{\rm diff}}^2$ is automatically divided into two parts. The first part $\sigma_0^2\propto S_{aa}(\omega)$ only relies on the auto-correlation $S_{aa}(\omega)$ of the noise and is the same as the 1-dim case~\cite{PhysRevResearch.3.023178, Wu:2024tcr}. The other part $\propto\sin2\theta\bar{S}_{a_xa_y}(\omega)$ characterises the contribution of the cross-correlation between $\delta a_x$ and $\delta a_y$ to the phase fluctuation. 

For $\theta=k\pi/2$ for some integer $k$, i.e. $\sin2\theta=0$, the cross-correlation term $\sin2\theta\bar{S}_{a_xa_y}(\omega_0)$ vanishes. In these cases, the $n$-axis is along the $x$- or $y$-direction, so the test mass is decoupled with the other direction, and the dephasing problem reduces into the 1-dimensional case. 

When the $n$-axis of the Bloch sphere of the internal state is not aligned to either $x$- or $y$-directions, the cross-term can contribute both positive and negative values to $\sigma_{\phi_{\rm diff}}^2$, which can be regarded as constructive and destructive interference of $\delta a_x$ and $\delta a_y$. Especially, $\sigma_{\phi_{\rm diff}}^2$ reaches its maximum and minimum value when $\sin2\theta=\pm1$, i.e. $\theta=\pi/4+k\pi$ for some integer $k$, in which case the $n$-axis is aligned to the diagonal lines of the $x$-$y$ plane. For the maximum destructive case, i.e. when $\sigma_{\phi_{\rm diff}}^2$ reaches its minimum value, the noise-induced dephasing is suppressed, which is a noise-mitigation strategy controlling internal states~\cite{PhysRevLett.132.223601}.

The maximum value of the cross-correlation term is constrained by the Cauchy-Schwarz inequality
\begin{equation}
    \left|\mathbb{E}[\delta a_x(t_1)\delta a_y(t_2)]\right|^2 \leq \mathbb{E}[\delta a_x(t_1)\delta a_x(t_2)]\mathbb{E}[\delta a_y(t_1)\delta a_y(t_2)],
\end{equation}
which implies an inequility $|S_{a_xa_y}(\omega_0)|\leq S_{aa}(\omega_0)$. Therefore, the phase variance $\sigma_{\phi_{\rm diff}}^2$ can never reach a negative value, which is physically plausible. In the next section, we will study the inertial force noise and optimise the parameters to minimise the variance $\sigma_{\phi_{\rm diff}}^2$.

\section{Dynamics of Inertial Noise}\label{section 4}

\begin{figure}
    \begin{tikzpicture}
        \coordinate (A) at (0, 0);
        \coordinate (B) at (5, 0);
        \coordinate (C) at (5, 4);
        \coordinate (D) at (0, 4);
        \filldraw[fill=cyan, fill opacity=0.02, draw=none] (A) -- (B) -- (C) -- (D) -- cycle;
        \draw[thick, line width=1.5pt] (A) -- (B) -- (C) -- (D) -- cycle;
        \node[font=\fontsize{11}{13}\selectfont] at (2.5, 4.25) {Experiment Box};

        \draw[->, blue!70, line width=1.25pt] (0.25, 0.25) -- (1.25, 0.25) node[black, right]{$x$};
        \draw[->, blue!70, line width=1.25pt] (0.25, 0.25) -- (0.25, 1.25) node[black, above]{$y$};

        \draw[<->, orange, line width=1pt] (1.8, -0.25) -- (3.2, -0.25) node[right]{$\delta \ddot{X}$};
        \draw[<->, orange, line width=1pt] (-0.25, 1.5) -- (-0.25, 2.5) node[above]{$\delta \ddot{Y}$};

        \draw[gray, fill=blue!60] (1.5, 1.5) circle (0.2);
        \draw[gray, fill=blue!15] (3.5, 2.5) circle (0.2);
        \draw[dashed] (1.5, 1.5) -- (3.5, 2.5);
    \end{tikzpicture}
    \caption{\small The concept of inertial noise. The test mass is set up inside an experimental apparatus which can be affected by vibration noises $\delta \ddot{X}$ and $\delta \ddot{Y}$. Consequently, the random motion of the experiment apparatus can lead to an inertial force $\delta a_x=-\delta\ddot{X}$ and $\delta a_y=-\delta\ddot{Y}$ on the test mass, resulting in a phase fluctuation of the interferometer. In this schematic diagram, we use capitalised $X$ and $Y$ for the motion of the experimental apparatus and use lowercase letters $x$ and $y$ for the test mass.}
    \label{intuition}
\end{figure}
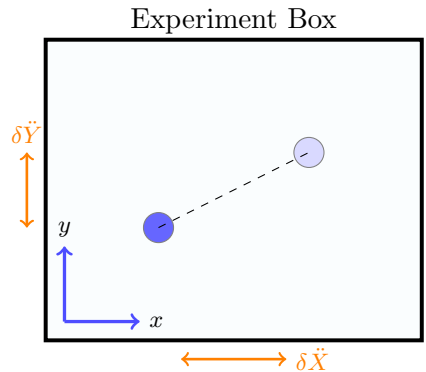

In this section, we will consider a typical source of the acceleration noise, often referred to as the inertial noise~\cite{PhysRevResearch.3.023178}, which is physically the inertial force noise on the test mass due to the random motion of the experiment apparatus~\footnote{We make a remark on the physical picture of the inertial noise. If the control and detection system can be fixed in the comoving reference frame of the test mass, then there is no stochastic inertial force exerted on the test mass. Consequently, there is no dephasing or spatial decoherence in the test mass. It intimates that decoherence of the test mass seems to be related to the reference frame. Some authors have studied some properties of quantum reference frames and also noticed similar effects~\cite{Giacomini2019}. For the classical stochastic reference frame case, further theoretical works are still required to understand the relevance of decoherence with respect to the reference frame.}.

As is illustrated as Fig.\ref{intuition}, the test mass is set up indside an experiment apparatus which is affected by some ambient noise like vibrations, denoted as $\delta \ddot{X}$ and $\delta \ddot{Y}$. Since the control and detection system should be supported by the the experimental box, the reference frame of the experiment can be naturally chosen as the comoving reference of the experiment apparatus. Consequently, the test mass will experience an inertial force noise satisfying 
\begin{equation}
    \delta a_x(t)=-\ddot{X},\quad \delta a_y(t)=-\ddot{Y},    
\end{equation}
where we use the capitalized letters representing the degrees of freedom (DOF) of the experiment apparatus and the lowercase letters for the DOFs of the interferometer. 

The motion of the experiment apparatus can be usually modelled as a 2-dimensional oscillator under dissipation and external stochastic forces, mathematically described by two independent Langevin equations or Ornstein-Uhlenbeck processes~\cite{Cai_2012, RevModPhys.17.323, PhysRevE.54.2084, 10.1119/1.18210}, 
\begin{equation}
\left\{\begin{aligned}
    \ddot{X} &= -\Omega_0^2X - \gamma_X\dot{X} + A_X(t), \\
    \ddot{Y} &= -\Omega_0^2Y - \gamma_Y\dot{Y} + A_Y(t),
\end{aligned}\right.
\end{equation}
where $\gamma_X$ and $\gamma_Y$ are dissipation rates, $\Omega_0$ is the intrinsic frequency of the experiment apparatus. $A_X(t)$ and $A_Y(t)$ are random noisy forces exerting on the experiment apparatus which can be modelled as independent white noises, of which the PSDs are constant in the frequency domain, i.e. $S_{A_iA_j}(\omega)=S_{A_iA_j}\delta_{ij}$ for $i, j=X,\ Y$, where $S_{A_iA_i}$ is the PSD of the white noise $A_i$. In this case, the vibrations along $X$- and $Y$-directions are decoupled, which indicates a zero cross-correlation between $X(t)$ and $Y(t)$. 

As we have discussed in the previous section, a non-zero cross-correlation of the 2-dimensional noise can suppress the dephasing effect by destructive interference, so one can introduce a coupling between $X(t)$ and $Y(t)$ to contribute a non-zero co-spectrum of $\delta a_x$ and $\delta a_y$. 

The most straightforward coupling between $X(t)$ and $Y(t)$ is the Coriolis force, which contributes a coupling described as a classical Hamiltonian
\begin{equation}
    H_{\rm rot} = 2\Omega_{\rm rot}(X\dot{Y}-Y\dot{X}).
\end{equation}
However, as shown in Appendix \ref{appendix b}, this coupling term contributes a pure-imaginary-valued cross-PSD of $\delta a_x$ and $\delta a_y$. Since the destructive interference requires a non-zero real part of the cross-PSD, the Coriolis force is not able to reduce the noises.

Therefore, we propose to phenomenologically introduce a classical Hamiltonian term 
\begin{equation}
    H_{\rm int} = kXY,
\end{equation}
to directly couple $X$ and $Y$~\footnote{The coupling term $H_{\rm int}$ can be very complicated in principle. However, the coupling term can be expanded as a Taylor series, of which the leading order term is exactly $H_{\rm int}=kXY$.}. This type of coupling term can be experimentally realised using devices made of elastic materials with a high shear modulus, which is often referred to as \emph{vibration direction converters}~\cite{Ito1972StudyOR, xu2018equivalent}. The dynamics of this kind of device are mathematically complicated, and we shall refrain from an extensive analysis of its dynamics. Under this coupling term, the 2-dimensional dynamical equations of the experimental apparatus are
\begin{equation}\label{EOM of apparatus}
\left\{\begin{aligned}
    \ddot{X} &= -\Omega_0^2X + kY - \gamma_X\dot{X} + A_X(t), \\
    \ddot{Y} &= -\Omega_0^2Y + kX - \gamma_Y\dot{Y} + A_Y(t),
\end{aligned}\right.
\end{equation}
For simplicity, we assume the dissipation force and the external random force are isotropic along $x$- and $y$-directions, i.e. $\gamma_X=\gamma_Y\equiv\gamma$ and $S_{A_xA_x}=S_{A_yA_y}\equiv S_0$. 

The normal modes of the oscillation of the experimental apparatus can be obtained by introducing
\begin{equation}
    U=\frac{X+Y}{\sqrt{2}},~~~V=\frac{X-Y}{\sqrt{2}}.
\end{equation}
In particular, the dynamical equation Eq.~\eqref{EOM of apparatus} implies the decoupled equations of $U$ and $V$,
\begin{equation}\label{EOM of normal mode}
\left\{\begin{aligned}
    \ddot{U} &= (-\Omega_0^2+k)U - \gamma\dot{U} + A_U(t), \\
    \ddot{V} &= (-\Omega_0^2-k)V - \gamma\dot{V} + A_V(t),    
\end{aligned}\right.
\end{equation}
where $A_U=(A_X+A_Y)/\sqrt{2}$ and $A_V=(A_X-A_Y)/\sqrt{2}$. Note that there is a constraint on the coupling term that is $|k|<\Omega_0^2$, otherwise either $U(t)$ or $V(t)$ diverges exponentially, resulting in divergence of $X(t)$ and $Y(t)$. 

\begin{figure*}
    \centering
    \begin{minipage}{0.3\textwidth}
        \includegraphics[scale=0.3]{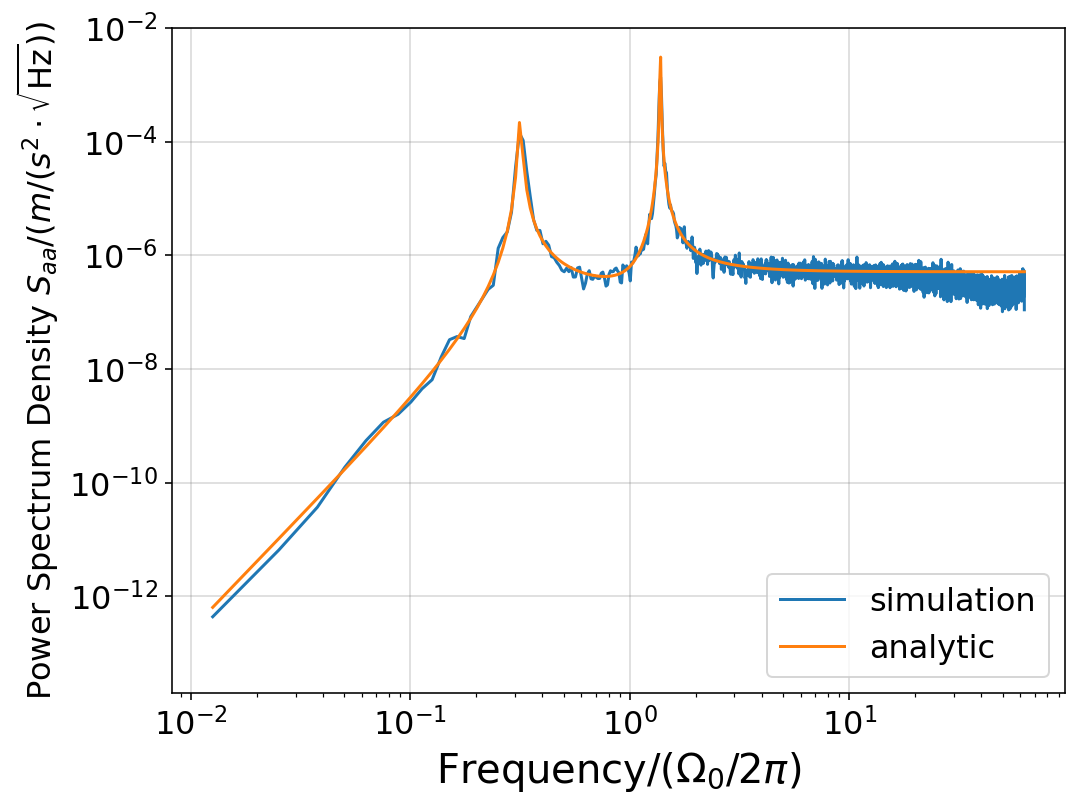}
        \subcaption{}
        \label{PSD a}
    \end{minipage}
    \hfill
    \begin{minipage}{0.3\textwidth}
        \includegraphics[scale=0.3]{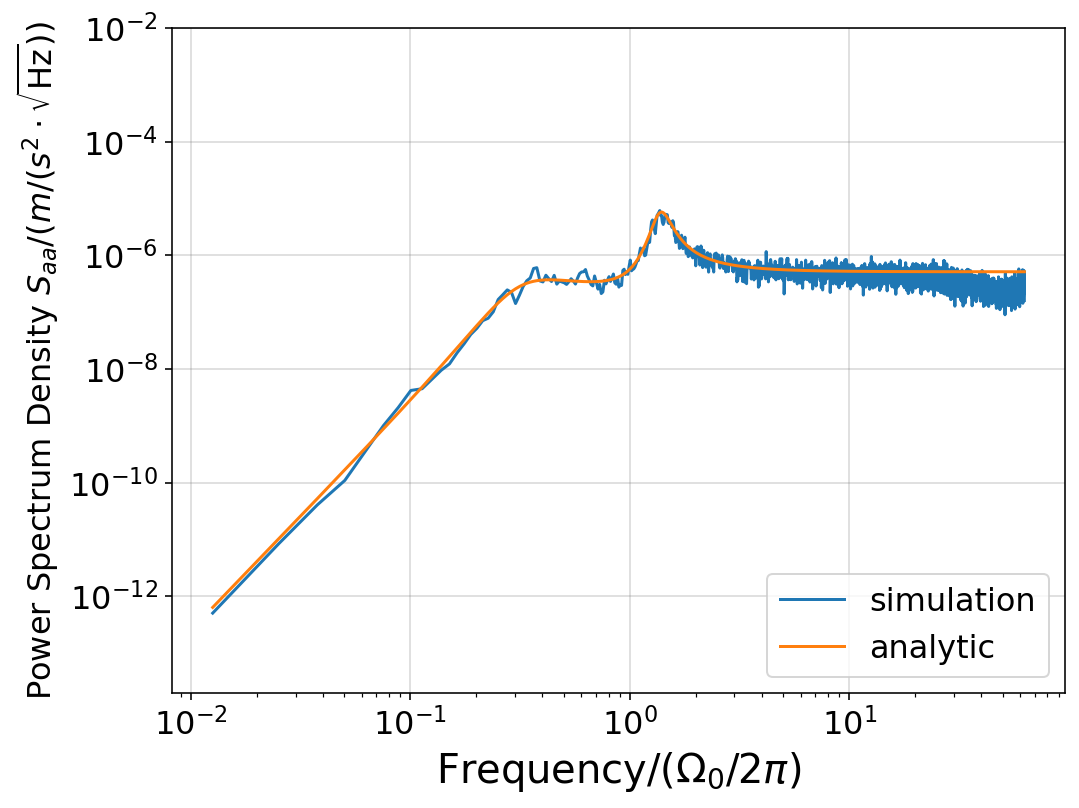}
        \subcaption{}
        \label{PSD b}
    \end{minipage}
    \hfill
    \begin{minipage}{0.3\textwidth}
        \includegraphics[scale=0.3]{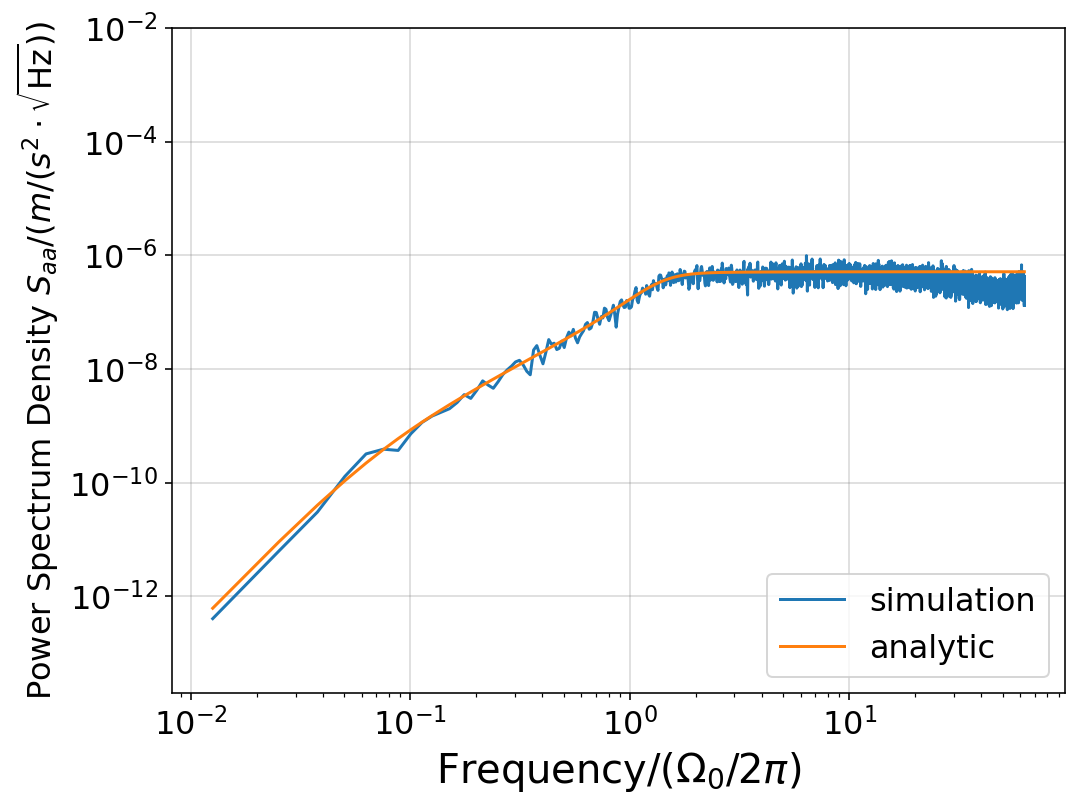}
        \subcaption{}
        \label{PSD c}
    \end{minipage}
    \hfill
    \begin{minipage}{0.3\textwidth}
        \includegraphics[scale=0.3]{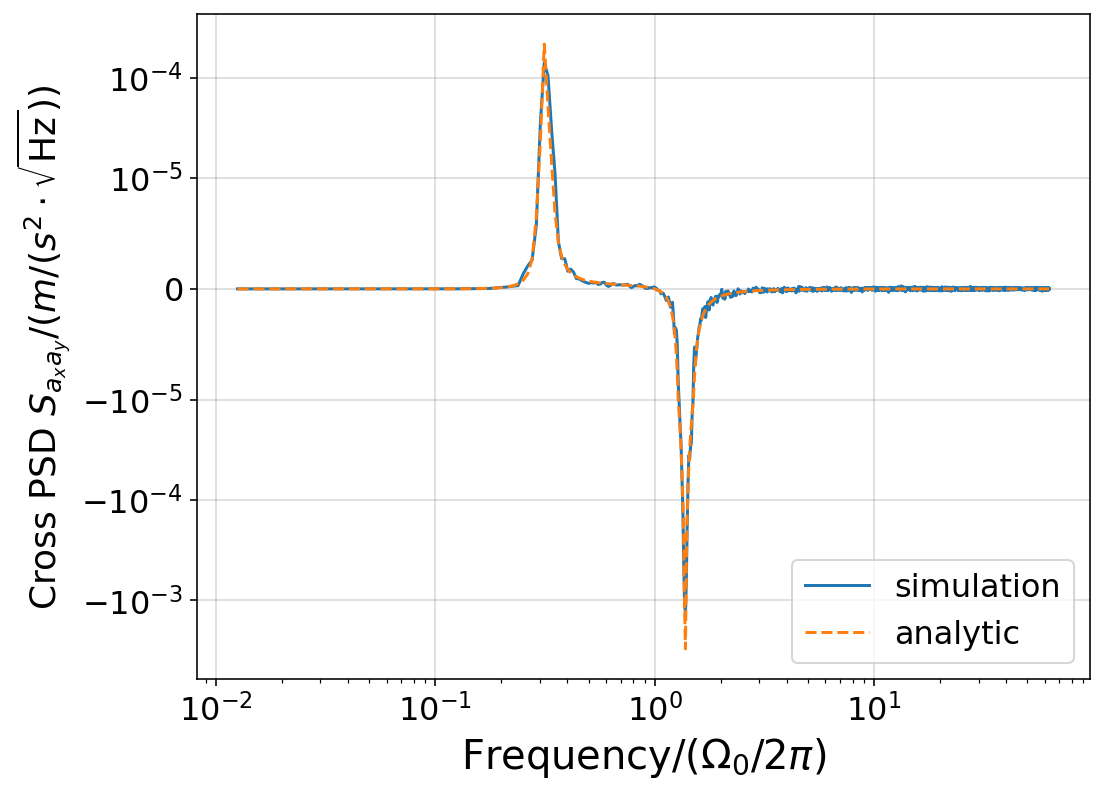}
        \subcaption{}
        \label{cross-PSD a}
    \end{minipage}
    \hfill
    \begin{minipage}{0.3\textwidth}
        \includegraphics[scale=0.3]{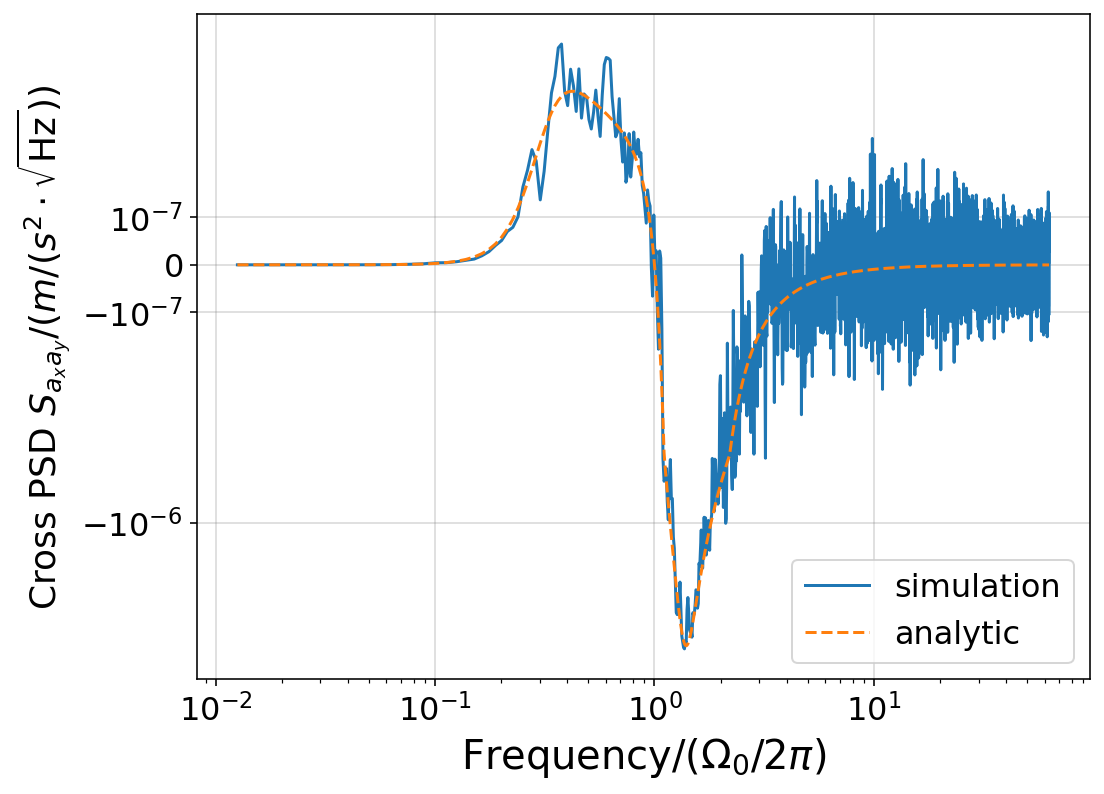}
        \subcaption{}
        \label{cross-PSD b}
    \end{minipage}
    \hfill
    \begin{minipage}{0.3\textwidth}
        \includegraphics[scale=0.3]{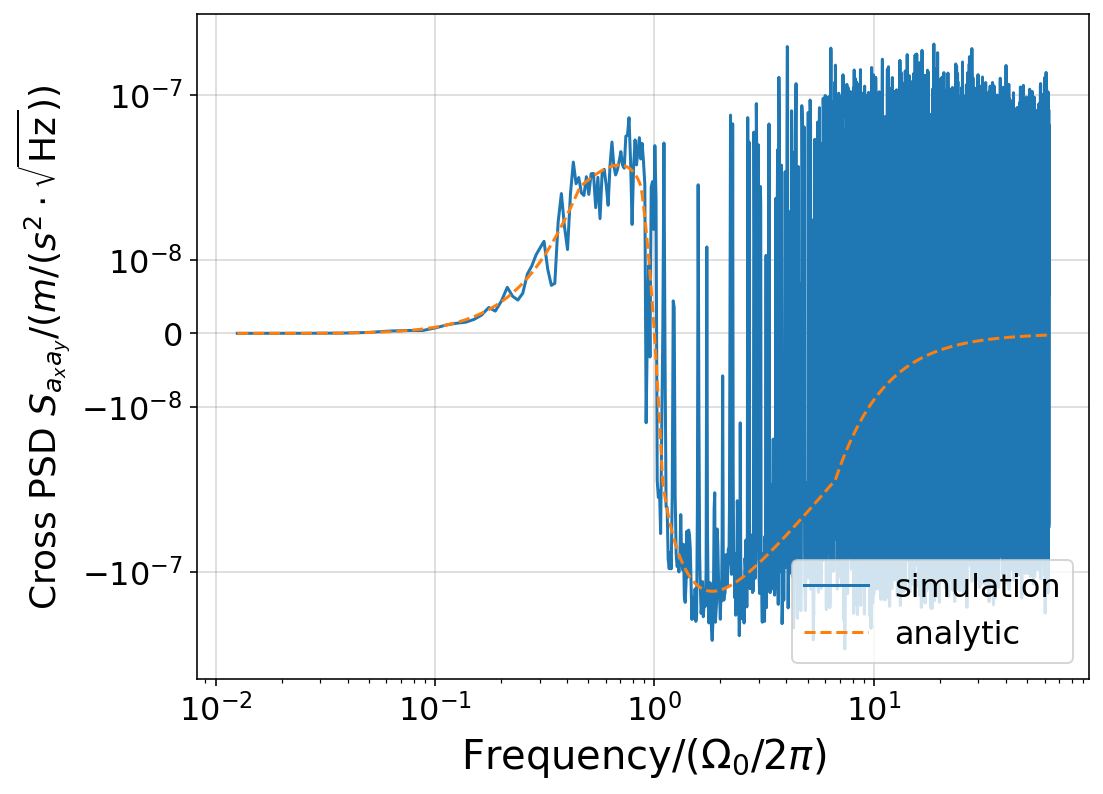}
        \subcaption{}
        \label{cross-PSD c}
    \end{minipage}
    \caption{\small The PSD $S_{aa}(\omega)$ and co-spectrum $\bar{S}_{a_xa_y}(\omega)$ with the parameters chosen as $k=0.9\Omega_0^2$ for all plots and $\gamma=0.01\,\Omega_0$, $\gamma=0.3\,\Omega_0$, $\gamma=1.5\,\Omega_0$ respectively. The asymptotic behavior of the PSD and the cross-PSD are $S_{aa}(\omega)\sim\omega^4$ and $\bar{S}_{a_xa_y}(\omega)\sim\omega^4$ in low frequency limit, and $S_{aa}(\omega)\rightarrow S_0$ and $\bar{S}_{a_xa_y}(\omega)\sim\omega^{-2}$ in high frequency limit. There are generally two resonance peaks of $S(\omega)$ at $\omega_{\rm peak}=\sqrt{\Omega_0^2\pm k-\gamma^2/4}$ with bandwidths $\Delta\omega=\gamma$, corresponding to the damped frequencies and the damping rate of the normal modes of the motion of the experiment apparatus. For $\bar{S}_{a_xa_y}(\omega)$ there are generally a positive resonance peak at $\omega_{\rm peak}=\sqrt{\Omega_0^2-k-\gamma^2/4}$ and a negative peak at$\omega_{\rm peak}=\sqrt{\Omega_0^2+k-\gamma^2/4}$ with bandwidths $\Delta\omega=\gamma$ for both peaks. For larger damping rate $\gamma$, the resonance peaks can be hidden due to the overdamped effect, shown as (b), (c), (e) and (f).}
    \label{PSD}
\end{figure*}

In the frequency space, the dynamical equations Eq.\,\eqref{EOM of apparatus} of the experiment apparatus are
\begin{equation}
\left\{\begin{aligned}
    -\omega^2X(\omega) = -\Omega_0^2X(\omega) + kY(\omega) - i\omega\gamma X(\omega) + A_X(\omega), \\
    -\omega^2Y(\omega) = -\Omega_0^2Y(\omega) + kX(\omega) - i\omega\gamma Y(\omega) + A_Y(\omega).
\end{aligned}\right.
\end{equation}
The solution can be written in a matrix form as
\begin{equation}\label{mechanical susceptibility}
    \begin{pmatrix}
        X(\omega) \\ Y(\omega)
    \end{pmatrix} = \frac{1}{\text{det}T}T\begin{pmatrix}
        A_X(\omega) \\ A_Y(\omega)
    \end{pmatrix}, \\
\end{equation}
where the transfer matrix $T$ and its determinant are
\begin{equation}
\begin{aligned}
    T &= \begin{pmatrix}
        \Omega_0^2-\omega^2+i\omega\gamma & k \\
        k & \Omega_0^2-\omega^2+i\omega\gamma
    \end{pmatrix}, \\
    \text{det}T &= \left(\Omega_0^2-\omega^2+i\omega\gamma\right)^2 - k^2.
\end{aligned}
\end{equation}
Note that the $T/\text{det}T$ is often referred to as the \emph{mechanical susceptibility}, see~\cite{bowen2015quantum} and is usually denoted by $\chi(\omega)$. Based on the susceptibility matrix, the PSD for $X$ and $Y$ and their correlations can be written in a matrix form as
\begin{widetext}
\begin{equation}
\begin{aligned}
    \begin{pmatrix}
        S_{XX}(\omega) & S_{XY}(\omega) \\
        S_{YX}(\omega) & S_{YY}(\omega)
    \end{pmatrix} &= S_0\chi(\omega)\chi^\dagger(\omega) 
        = \frac{S_0}{|\text{det}T|^2} \begin{pmatrix}
        (\Omega_0^2-\omega^2)^2+\omega^2\gamma^2+k^2 & 2k(\Omega_0^2-\omega^2) \\
        2k(\Omega_0^2-\omega^2) & (\Omega_0^2-\omega^2)^2+\omega^2\gamma^2+k^2
    \end{pmatrix}, \\ \\
    |\text{det}T|^2 &= \left((\Omega_0^2-k-\omega^2)^2+\omega^2\gamma^2\right) \left((\Omega_0^2+k-\omega^2)^2+\omega^2\gamma^2\right).
\end{aligned}
\end{equation}
\end{widetext}
Since $a_x(t)=-\ddot{X}$ and $a_y(t)=-\ddot{Y}$, then the PSD $S_{aa}(\omega)$ and the cross PSD $S_{a_xa_y}(\omega)$ satisfy $S_{aa}(\omega)=\omega^4S_{XX}(\omega)$ and $S_{a_xa_y}(\omega)=\omega^4S_{XY}(\omega)$, i.e.
\begin{equation}\label{Saa and Saxay}
\begin{aligned}
    S_{aa}(\omega) &= \omega^4S_0\frac{(\Omega_0^2-\omega^2)^2+\omega^2\gamma^2+k^2}{|\text{det}T|^2}, \\
    S_{a_xa_y}(\omega) &=\omega^4S_0\frac{2k(\Omega_0^2-\omega^2)}{|\text{det}T|^2}.
\end{aligned}
\end{equation}
Fig.\,\ref{PSD} show the analytical result and a simulation of the PSD $S_{aa}(\omega)$ and the co-spectrum $S_{a_xa_y}(\omega)$ under different parameters. The parameters are chosen as $k=0.9\Omega_0^2$ for all plots, while the damping rate $\gamma$ is chosen as $0.01\,\Omega_0$, $0.3\,\Omega_0$ and $1.5\,\Omega_0$, corresponding to typical scenarios of the underdamped case, the overdamped one-mode case, and the overdamped two-mode case, respectively. Further details are discussed below.

This simulation is based on the 4th-order Runge-Kutta method. We first simulated the motion of $X(t)$ and $Y(t)$, then used the Fast Fourier Transform (FFT) to compute the corresponding PSD and cross-PSD. More technical details of the simulation are summarised in Appendix \ref{appendix d}.

Generally speaking, $S_{aa}(\omega)$ is proportional to $\omega^4$ in low frequency, and it tends to a constant $S_0$ in high frequency, as shown in Fig.\,\ref{PSD}. Besides, $S_{a_xa_y}(\omega)$ is also proportional to $\omega^4$ in the low-frequency limit, but it decreases as a speed $\omega^{-2}$, shown as the last 3 subfigures of Fig.\,\ref{PSD}.

In addition, there are generally two resonance peaks of $S_{aa}(\omega)$ and $S_{a_xa_y}(\omega)$, corresponding to the normal modes of the dynamical equation Eq.\,\eqref{EOM of apparatus} which can be regarded as two coupled oscillators. Note that the left peak and right peak of $S_{a_xa_y}(\omega)$ are positive and negative respectively, due to the factor $(\Omega_0^2-\omega^2)$ in Eq.\,\eqref{Saa and Saxay}, shown as Fig.\,\ref{cross-PSD a}. The normal mode frequencies are exactly the poles of the mechanical susceptibility matrix $\chi(\omega)$, which can be determined by solving the equation $\text{det}T=0$. The analytic result for the poles of $\chi(\omega)$ are given by
\begin{equation}
\begin{aligned}
    \omega_{1,2} &= \frac{i}{2}\gamma \pm \frac{1}{2}\sqrt{4(\Omega_0^2-k)-\gamma^2}, \\
    \omega_{3,4} &= \frac{i}{2}\gamma \pm \frac{1}{2}\sqrt{4(\Omega_0^2+k)-\gamma^2}.
\end{aligned}
\end{equation}
The real and imaginary parts of $\omega_{1-4}$ corresponds to the resonance peak positions and widths of $S_{aa}(\omega)$, so the two positive peaks locate at $\omega_{1,3}=\sqrt{\Omega_0^2\pm k-\gamma^2/4}$. Note that these two peak frequencies are exactly the damped frequencies of the normal modes formulated in the dynamical equations Eq.\,\eqref{EOM of normal mode}. 

The sharpness of the peaks is usually characterised by the Q-factors, which are usually defined by the ratio between the peak frequencies and the bandwidths of peaks, i.e. 
\begin{equation}\label{Q-factor}
    Q\equiv \frac{\omega_{\rm peak}}{\Delta\omega}.
\end{equation}
Note that the imaginary part of the poles $\omega_{1-4}$ represents the bandwidths of the peaks, so the Q-factors of the two positive peaks are
\begin{equation}
    Q_{1,3} = \frac{\sqrt{\Omega_0^2\pm k-\gamma^2/4}}{\gamma}.
\end{equation}
For a small damping rate $\gamma\ll\sqrt{\Omega_0^2\pm k}$, the damping rate can reduce the oscillation frequencies of the normal modes by $\gamma^2/(8\sqrt{\Omega_0^2\pm k})$. On the other hand, if the damping rate is so large that $\gamma^2>\Omega_0^2\pm k$, then the normal modes cease to oscillate and instead exhibit exponential decay, evidenced by the disappearance of peaks in $S_{aa}(\omega)$, known as the overdamping effect. For an intermediate damping rate $\gamma\in(\sqrt{\Omega_0^2-k},\,\sqrt{\Omega_0^2+k})$, there is only one peak at $\omega=\sqrt{\Omega_0^2+k-\gamma^2/4}$, as shown in Fig.\,\ref{PSD b} and Fig.\,\ref{cross-PSD b}. For a large damping rate $\gamma>\sqrt{\Omega_0^2+k}$, there is no resonance peak, as shown in Fig.\,\ref{PSD c} and Fig.\,\ref{cross-PSD c}.

\section{Inertial Noise Induced Dephasing}

Based on the PSD and the cross-PSD \eqref{Saa and Saxay}, and noticing that $S_{aa}(\omega\to0)=0$ and $S_{a_xa_y}(\omega\to0)=0$, the variance $\sigma_{\phi_{\rm diff}}^2$ of the phase fluctuation \eqref{sigma simplified} is given by:
\begin{equation}
\begin{split}
    \sigma_{\phi_{\rm diff}}^2 = \frac{8\pi^2m^2A_0^2}{\hbar^2}S_0 &\bigg[ \frac{\omega_0^3 \cos^2\left(\theta+\frac{\pi}{4}\right)}{(\Omega_0^2+k-\omega_0^2)^2+\omega_0^2\gamma^2} \\
        +& \frac{\omega_0^3 \sin^2\left(\theta+\frac{\pi}{4}\right)}{(\Omega_0^2-k-\omega_0^2)^2+\omega_0^2\gamma^2} \bigg].
\end{split}
\end{equation}
It is noteworthy that the two Lorentzian distributions exactly correspond to the PSDs of the normal modes $U=(X+Y)/\sqrt{2}$ and $V=(X-Y)/\sqrt{2}$ of the noise. In addition, the factors $\cos(\theta+\pi/4)$ and $\sin(\theta+\pi/4)$ represent the projection of the direction of the superposition on the directions of the normal modes $U$ and $V$. Thus, the phase variance $\sigma_{\phi_{\rm diff}}^2$ can be exactly written as
\begin{equation}
\begin{split}
    \sigma_{\phi_{\rm diff}}^2 = \frac{8\pi^2m^2A_0^2}{\hbar^2}S_0\omega_0^3 &\bigg[S_{VV}(\omega_0)\cos^2\left(\theta+\frac{\pi}{4}\right) \\
        +& S_{UU}(\omega_0)\sin^2\left(\theta+\frac{\pi}{4}\right)\bigg].
\end{split}
\end{equation}
In other words, the effect of the vibration coupling term $H_{\rm int}=kXY$ on the dephasing of the test mass is that $\sigma_{\phi_{\rm diff}}^2$ resonates to $U$ and $V$ modes instead of the initial vibration modes $X$ and $Y$.

For a positive coupling $k$, it is not difficult to verify that $S_{VV}(\omega_0)<S_{UU}(\omega_0)$ when $\omega_0<\Omega_0$, then the variance $\sigma_{\phi_{\rm diff}}^2$ reaches its minimum value $\propto S_{VV}(\omega_0)$ when $\cos^2(\theta+\pi/4)=1$ and $\sin^2(\theta+\pi/4)=0$. By contrast, when $\omega_0>\Omega_0$, $\sigma_{\phi_{\rm diff}}^2$ reaches its minimum value $\propto S_{UU}(\omega_0)$ when $\cos^2(\theta+\pi/4)=0$ and $\sin^2(\theta+\pi/4)=1$.

Remarkably, if the coupling $k$ is allowed both positive and negative, then $S_{VV}(\omega_0)$ and $S_{UU}(\omega_0)$ have the same form under a transform $k\to-k$. Thus, the minimum value of $\sigma_{\phi_{\rm diff}}^2$ always has a form given by:
\begin{equation}
    \sigma_{\phi_{\rm diff}}^2 = \frac{8\pi^2m^2A_0^2}{\hbar^2}S_0\frac{\omega_0^3}{(\Omega_0^2-k-\omega_0^2)^2+\omega_0^2\gamma^2},
\end{equation}
where $k>0$ for $\omega_0>\Omega_0$ and $k<0$ for $\omega_0<\Omega_0$. Notably, compared to the dephasing induced by a 1-dim noise, the 2-dim case is almost the same except for a resonance peak translation $\Omega_0^2\to\Omega_0^2-k$.

There are still two free parameters $\gamma$ and $k$, where $\omega_0$ and $\Omega_0$ are given by the experimental conditions. Then the dephasing can be further minimised by optimising $\gamma$ and $k$, or equivalently, the resonance peak frequency and the corresponding Q-factor.

\begin{figure*}
    \centering
    \begin{minipage}{0.45\textwidth}
        \includegraphics[scale=0.4]{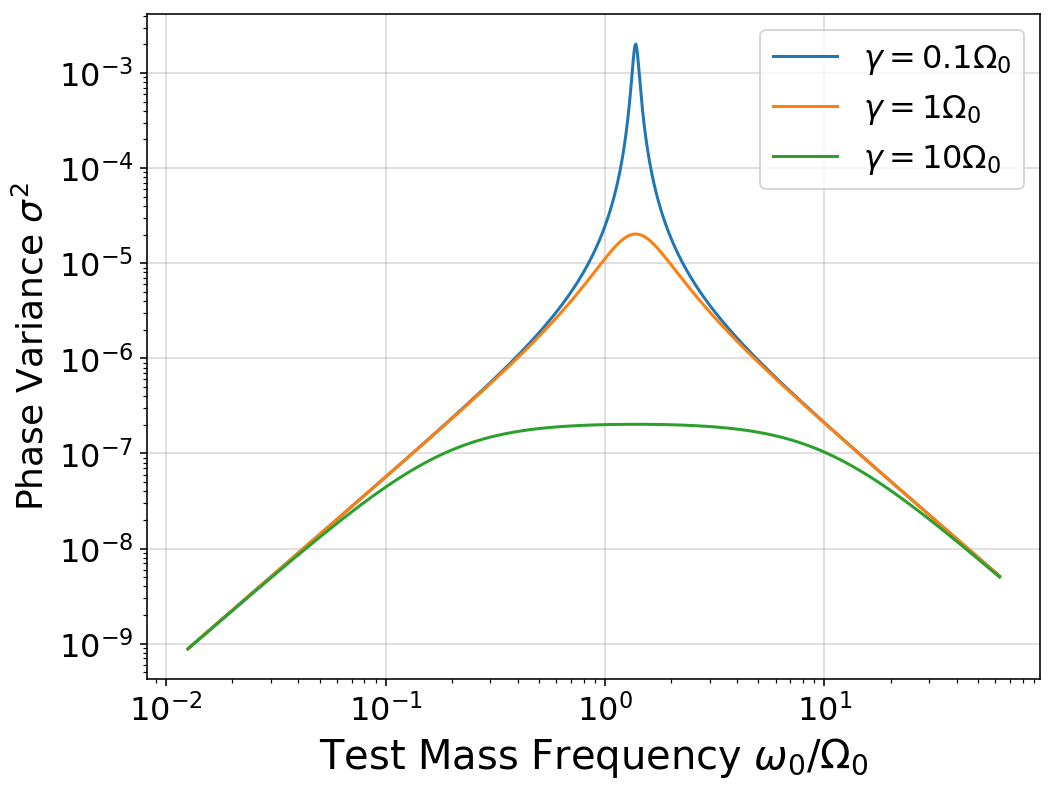}
        \subcaption{\small $k=0.9\,\Omega_0^2$.}
        \label{sigma gamma}
    \end{minipage}
    \hfill
    \begin{minipage}{0.45\textwidth}
        \includegraphics[scale=0.4]{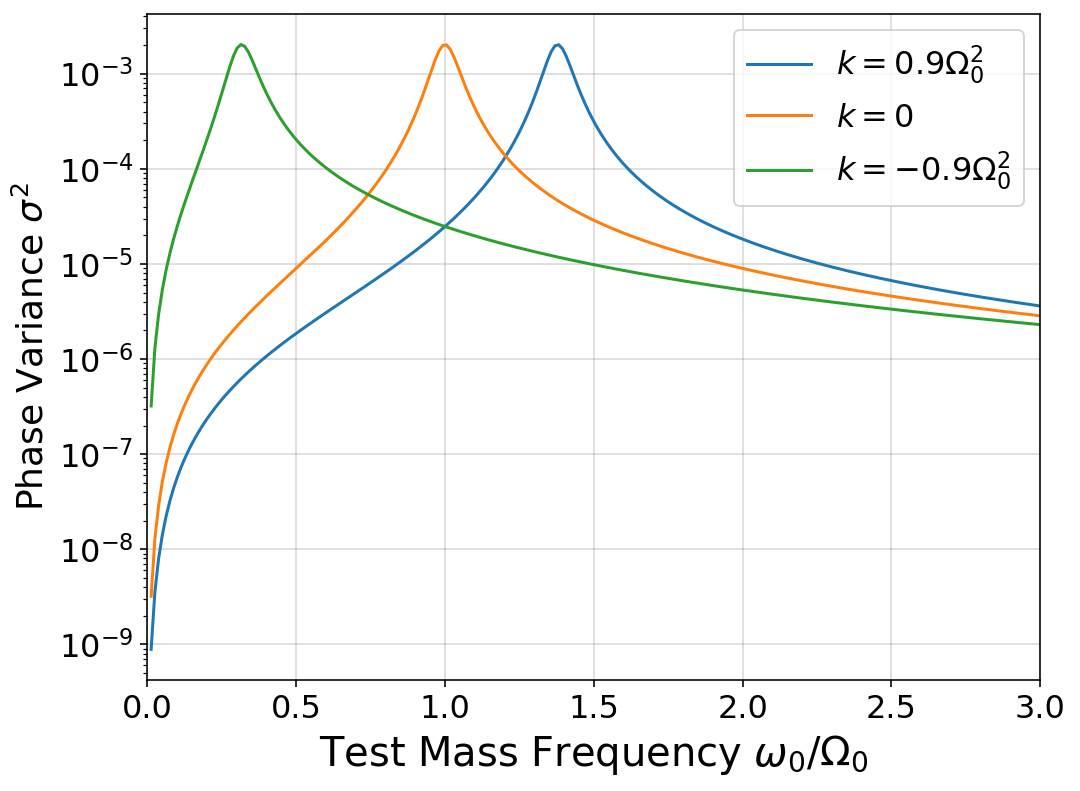}
        \subcaption{\small $\gamma=0.1\,\Omega_0$.}
        \label{sigma k}
    \end{minipage}
    \caption{\small $\sigma^2_{\phi_{\rm diff}}$ under different parameters. In subfigure (a), the coupling $k$ is fixed $0.9\,\Omega_0^2$, while the damping rate $\gamma$ varies from $10^{-1}\,\Omega_0$ to $10\,\Omega_0$. As is shown, the damping rate $\gamma$ can change the height and width of the peak, but it doesn't affect $\sigma_{\phi_{\rm diff}}^2$ when $\omega_0$ is far from $\Omega_0$. In subfigure (b), the damping rate is fixed as $0.1\,\Omega_0$, while the coupling $k$ varies from $-0.9\,\Omega_0$ to $0.9\,\Omega_0$. As is shown, the coupling $k$ doesn't change the shape of the peak, but it can translate the peak position. When $\omega_0$ nearly resonates to $\Omega_0$, $k$ can significantly suppress $\sigma_{\phi_{\rm diff}}^2$, approximately characterized by $Q^2$.}
    \label{sigma}
\end{figure*}

Fig.\,\ref{sigma} illustrates the relationship between the dephasing factor $\sigma_{\phi_{\rm diff}}^2$ and the intrinsic frequency $\omega_0$ under varying parameters. Specifically, Fig.\,\ref{sigma gamma} presents the result with a fixed coupling $k$, while Fig.\,\ref{sigma k} shows the result with a fixed damping rate $\gamma$. Both plots demonstrate that the asymptotic behaviour of $\sigma_{\phi_{\rm diff}}^2$ with respect to $\omega_0$ is that $\sigma_{\phi_{\rm diff}}^2$ is proportional to $\omega_0$ in small $\omega_0$ limit and to $\omega_0^{-1}$ in large $\omega_0$ limit.

Fig.\,\ref{sigma gamma} shows $\sigma_{\phi_{\rm diff}}$ under different damping rate $\gamma$ with a certain coupling $k=0.9\,\Omega_0^2$. As is shown, the damping rate can suppress the height of the peak with a ratio $(\gamma/\Omega_0)^2$, which is exactly $Q^2$. On the other hand, $\gamma$ is not able to influence $\sigma_{\phi_{\rm diff}}^2$ in small and large $\omega_0$ limits. So if $\omega_0$ is far from $\Omega_0$, it is useless to change $\gamma$.   

Fig.\,\ref{sigma k} shows $\sigma_{\phi_{\rm diff}}$ under different coupling $k$ with a fixed damping rate $\gamma=0.1\,\Omega_0$. As is shown, the coupling $k$ doesn't change the peak value but only translates the peak position from $\Omega_0$ to $\sqrt{\Omega_0^2+k}$. Consiquently, $k$ can slightly change the value of $\sigma_{\phi_{\rm diff}}^2$ in the small $\omega_0$ limit. In addtion, if $\omega_0$ resonantes with the noise frequency $\Omega_0$, $k$ can significantly reduce $\sigma_{\phi_{\rm diff}}$ approximately from $\propto 1/(\omega_0^2\gamma^2)$ to $\propto 1/(k^2+\omega_0^2\gamma^2)$. Using the relation $\omega_0\sim\Omega_0$ and $k\sim\Omega_0^2$, the suppression ratio is approximately $Q^2$, see Eq.(\ref{Q-factor}). 

\section{Discussion}\label{section 6}

In the end, we draw some conclusions on the application of noise mitigation in the gravity experiment. For a gravity experiment based on matter-wave interferometers, the gravitational interaction can couple to the superposition of the test mass, and then the differential phase of the two arms at the final time encodes the information about gravity.

As an example, we would like to discuss the gravimeter based on nitrogen-vacancy (NV) centre driven by magnetic field~\cite{PhysRevLett.111.180403, Zhang2018}. In this experiment, the gravitational acceleration of the Earth can lead to an interaction term on the interferometer by $\hat{H}_{\rm grav}=mg\hat{x}$ when the direction of $x$-axis is chosen as the vertical direction. Then this coupling term can lead to a phase difference between the two arms as
\begin{equation}\label{diff phi}
    \phi_{\rm diff} = \frac{1}{\hbar}\int mg\left(x_+(t)-x_-(t)\right)\,\mathrm{d}t.
\end{equation}
When the ideal phase along the trajectories follows \eqref{ideal trajectories}, this signal differential is formulated by
\begin{equation}
    \phi_{\rm diff} = \frac{2\pi mgA_0\cos\theta}{\hbar\omega_0}.
\end{equation}
Here, the angle $\theta$ defined by \eqref{angle theta} represents the angle between the $n$-axis of the Bloch sphere and the direction of gravitational acceleration. Then one can compute the signal-to-noise ratio
\begin{equation}
    \text{SNR} \equiv \frac{\phi_{\rm diff}}{\sigma_{\phi_{\rm diff}}} = \frac{g\cos\theta}{\omega_0^2}\sqrt{\frac{(\Omega_0^2-k-\omega_0^2)^2+\omega_0^2\gamma^2}{2S_0\omega_0}}.
\end{equation}
As discussed in the previous section, when the noise approximately resonates with the test mass (i.e. $\omega_0\approx\Omega_0$), the variance $\sigma_{\phi_{\rm diff}}^2$ can be suppressed by a factor approximated to $Q^2$ under appropriate coupling $k$. Therefore, in this scenario, the SNR can be enhanced by a factor approximately equal to the $Q$-factor of the noise, i.e. see Eq.(\ref{Q-factor}). This example illustrates how cross correlation between the noises along
different directions, resulting in the suppression of the
dephasing of the interferometer. 

Hopefully, the results of our paper can be translated to two adjacent interferometers. The current analysis needs to be revisited to investigate how the cross correlations between interferometers can help reduce inertial dephasing between the common modes of the two interferometers. We will leave this analysis for future studies. This strategy will be particularly beneficial for testing entanglement, given the quantum nature of matter and gravity.

\begin{acknowledgements}
    M. Wu would like to thank the China Scholarship Council (CSC) for financial support.
    MT acknowledges funding from the Slovenian Research and Innovation Agency (ARIS) under contracts N1-0392, P1-0416, SN-ZRD/22-27/0510 (RSUL Toro\v{s}).
    AM's research is funded in part by the Gordon and Betty Moore Foundation through Grant GBMF12328, DOI 10.37807/GBMF12328. AM's work is supported by the Alfred P. Sloan Foundation under Grant No. G-2023-21130. S.B. would like to acknowledge EPSRC Grants No. EP/N031105/1, No. EP/S000267/1, and No. EP/X009467/1 and STFC Grant No. ST/W006227/1.

\end{acknowledgements}

\appendix
\section{Proof of \eqref{sigma simplified}}\label{appendix c}

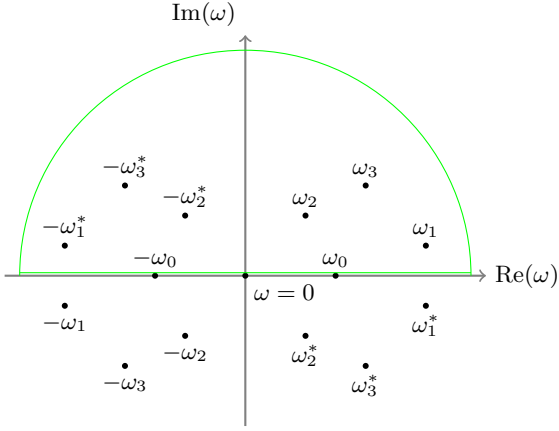
\begin{figure}
    \centering
    \begin{tikzpicture}[scale=0.8]
        \draw[->, thick, gray] (-4,0)--(4,0) node[anchor=west, black] {Re($\omega$)};
        \draw[->, thick, gray] (0,-2.5)--(0,4) node[anchor=south east, black] {Im($\omega$)};

        \fill (0,0) circle (0.05) node[anchor=north west] {$\omega=0$};
        \fill (1.5,0) circle (0.05) node[anchor=south] {$\omega_0$}; 
        \fill (-1.5,0) circle (0.05) node[anchor=south] {$-\omega_0$};
        \fill (3,0.5) circle (0.05) node[anchor=south] {$\omega_1$}; 
        \fill (3,-0.5) circle (0.05) node[anchor=north] {$\omega_1^*$};
        \fill (-3,0.5) circle (0.05) node[anchor=south] {$-\omega_1^*$}; 
        \fill (-3,-0.5) circle (0.05) node[anchor=north] {$-\omega_1$};
        \fill (1,1) circle (0.05) node[anchor=south] {$\omega_2$}; 
        \fill (1,-1) circle (0.05) node[anchor=north] {$\omega_2^*$};
        \fill (-1,1) circle (0.05) node[anchor=south] {$-\omega_2^*$}; 
        \fill (-1,-1) circle (0.05) node[anchor=north] {$-\omega_2$};
        \fill (2,1.5) circle (0.05) node[anchor=south] {$\omega_3$}; 
        \fill (2,-1.5) circle (0.05) node[anchor=north] {$\omega_3^*$};
        \fill (-2,1.5) circle (0.05) node[anchor=south] {$-\omega_3^*$}; 
        \fill (-2,-1.5) circle (0.05) node[anchor=north] {$-\omega_3$};
        
        \draw[green] (-3.75,0.05)--(3.75,0.05);
        \draw[green] (3.75,0) arc (0:180:3.75);
    \end{tikzpicture}
    \caption{\small Poles of the integrand in \eqref{integral I} and the integral path in the complex plane. The poles $\pm\omega_0$ arise from the transfer function $F_0(\omega)$, while other poles arise from the PSD and the co-spectrum.}
    \label{poles of integrand}
\end{figure}

In this appendix, we will prove that the integral \eqref{sigma integral} is equal to \eqref{sigma simplified} by the residue theorem. First of all, we rewrite the transfer function $F_0(\omega)$ as the real part of a complex function as 
\begin{equation}
    F_0(\omega) = 2\omega_0^4\text{Re}\left[\frac{1-\mathrm{e}^{2\pi i\omega/\omega_0}}{\omega^2(\omega^2-\omega_0^2)^2}\right].
\end{equation}
Besides, we denote $S(\omega)=S_{aa}(\omega)+\sin2\theta \bar{S}_{a_xa_y}(\omega)$, then the integral in \eqref{sigma integral} is exactly the real part of the integral
\begin{equation}\label{integral I}
    I = 2\omega_0^4\int S(\omega)\frac{1-\mathrm{e}^{2\pi i\omega/\omega_0}}{\omega^2(\omega^2-\omega_0^2)^2}\,\mathrm{d}\omega \equiv 2\omega_0^4\int f(\omega)\,\mathrm{d}\omega,
\end{equation}
where we denote the integrand in \eqref{integral I} as $f(\omega)\equiv S(\omega)(1-\mathrm{e}^{2\pi i\omega/\omega_0})/\omega^2(\omega^2-\omega_0^2)^2$. Then the poles of $f(\omega)$ consist of $0$, $\pm\omega_0$ and the poles of $S(\omega)$. 
Noticing that $S(\omega)$ is a real-valued even function on the real axis, we state that the poles of $S(\omega)$ are symmetric to both the real axis and the imaginary axis, shown as Fig.\ref{poles of integrand}. The proof of this statement is summarised at the end of this appendix.

According to the residue theorem, this integral equals the sum of the residues in the upper half of the complex plane, including those on the real axis. As is shown in Fig.\,\ref{poles of integrand}, $\omega_j$ and $-\omega^*_j$ are in the upper half of the complex plane, while $0$ and $\pm\omega_0$ are located on the real axis, so the integral \eqref{integral I} equals to
\begin{equation}
\begin{split}
    I/(2\omega_0^4) &= 2\pi i\left[\sum\limits_{j}\left(\mathop{\text{Res}}\limits_{\omega=\omega_j}f(\omega) + \mathop{\text{Res}}\limits_{\omega=-\omega_j^*}f(\omega)\right)\right] \\
        &+ \pi i \left(\mathop{\text{Res}}\limits_{\omega=0}f(\omega) + \mathop{\text{Res}}\limits_{\omega=\omega_0}f(\omega) + \mathop{\text{Res}}\limits_{\omega=-\omega_0}f(\omega)\right).
\end{split}
\end{equation}

In order to evaluate the constribution of the poles $\omega_j$ and $-\omega_j^*$, we firstly verify that $f(-\omega^*)=[f(\omega)]^*$. In particular, one can directly write
\begin{equation}
    f(-\omega^*) = S(-\omega^*)\frac{1-\mathrm{e}^{-2\pi i\omega^*/\omega_0}}{\omega^{*2}(\omega^{*2}-\omega_0^2)^2}.
\end{equation}
It is obvious that $1-\mathrm{e}^{-2\pi i\omega^*/\omega_0}=(1-\mathrm{e}^{2\pi i\omega/\omega_0})^*$ and $\omega^{*2}(\omega^{*2}-\omega_0^2)^2=[\omega^2(\omega^2-\omega_0^2)^2]^*$. Besides, according to the properties $S(-\omega)=S(\omega)$ and $S(\omega^*)=[S(\omega)]^*$, one can obtain $S(-\omega^*)=[S(\omega)]^*$. Combine these equations, one may find $f(-\omega^*)=[f(\omega)]^*$. 

Based on this property, the residue of $f(\omega)$ at the pole $-\omega_j^*$ satisfies 
\begin{equation}
    \mathop{\text{Res}}\limits_{\omega=-\omega_j^*}f(\omega) = \mathop{\text{Res}}\limits_{\omega=-\omega_j^*}\left[f(-\omega^*)\right]^* = \left[\mathop{\text{Res}}\limits_{\omega=\omega_j}f(\omega)\right]^*.
\end{equation}
Therefore, $2\pi i\left(\mathop{\text{Res}}\limits_{\omega=\omega_j}f(\omega)+\mathop{\text{Res}}\limits_{\omega=-\omega_j^*}f(\omega)\right)$ is exactly a pure imaginary number. As a consequence, the summation of all the poles $\omega_j$ and $-\omega_j^*$ doesn't contribute to $\sigma^2_{\phi_{\rm diff}}\propto\text{Re}I$. In other words, $\text{Re}I$ can be completely determined by the poles $\omega=0$ and $\pm\omega_0$.

According to the symmetry of $f(\omega)$, the residue values of the poles $\pm\omega_0$ are equal. Since both of them are 2nd-order poles, their contribution to the integral \eqref{integral I} is
\begin{equation}
\begin{split}
    2\pi i\mathop{\text{Res}}\limits_{\omega=\omega_0}f(\omega) &= 2\pi i\frac{\mathrm{d}}{\mathrm{d}\omega}\left[S(\omega)\frac{1-\mathrm{e}^{2\pi i\omega/\omega_0}}{\omega^2(\omega+\omega_0)^2}\right]\bigg|_{\omega=\omega_0} \\
        &= \frac{\pi^2}{\omega_0^5}S(\omega_0). 
\end{split}
\end{equation}

As for the other 2nd-order pole $\omega=0$, it contributes to the integral \eqref{integral I} as
\begin{equation}
\begin{split}
    \pi i\mathop{\text{Res}}\limits_{\omega=0}f(\omega) &= \pi i\frac{\mathrm{d}}{\mathrm{d}\omega}\left[S(\omega)\frac{1-\mathrm{e}^{2\pi i\omega/\omega_0}}{(\omega^2-\omega_0^2)^2}\right]\bigg|_{\omega=0} \\
        &= \frac{2\pi^2}{\omega_0^5}S(\omega\to0).
\end{split}
\end{equation}

In summary, the real part of the integral \eqref{integral I} is
\begin{equation}
    \text{Re}I = \frac{2\pi^2}{\omega_0}\left[S(\omega_0)+2S(\omega\to0)\right]. 
\end{equation}

After the proof of \eqref{sigma simplified}, we remark that the transfer function $F_0(\omega)$ mathematically behaves like a sum of delta-functions $4\pi^2\left[\delta(\omega-\omega_0)+2\delta(\omega)\right]/\omega_0$, although its shape Fig.\,\ref{transfer} is different from the delta-function. In fact, by introducing a rectangle function 
\begin{equation}
    \text{rect}(t) \equiv \left\{\begin{aligned}
        &1,\quad 0\leq t\leq\frac{2\pi}{\omega_0}, \\
        &0,\quad \text{others},
    \end{aligned}\right.
\end{equation}
the integral in the definition \eqref{F0} of the $F_0(\omega)$ can be written as 
\begin{equation}
    \int_{-\infty}^{\infty}(1-\cos\omega_0t)\text{rect}(t)\mathrm{e}^{i\omega t}\,\mathrm{d}t.
\end{equation}
This integral is exactly the Fourier transform of the function $(1-\cos\omega_0t)\text{rect}(t)$. According to the convolution theorem, this integral equals to the convolution of the Fourier transforms of these two functions, i.e. $\mathcal{F}(1-\cos\omega_0t)\ast\mathcal{F}(\text{rect}(t))/2\pi$, where these two Fourier transforms are given by $\mathcal{F}(1-\cos\omega_0t)=2\pi\left[\delta(\omega)-\delta(\omega-\omega_0)/2-\delta(\omega+\omega_0)/2\right]$ and $\mathcal{F}(\text{rect}(t))=(2\pi/\omega_0)\mathrm{e}^{\pi i\omega/\omega_0}\text{sinc}(\pi\omega/\omega_0)$. Therefore, $F_0(\omega)$ is exactly the convolution between the delta functions and the sinc-function
\begin{equation}
\begin{split}
    F_0(\omega) = &\frac{\pi^2}{\omega_0^2}\bigg| \left[2\delta(\omega)-\delta(\omega-\omega_0)/2-\delta(\omega+\omega_0)/2\right] \\
        &\ast \left[\mathrm{e}^{\pi i\frac{\omega}{\omega_0}}\text{sinc}\frac{\pi\omega}{\omega_0}\right] \bigg|^2.
\end{split}
\end{equation}
Hence, $F_0(\omega)$ has a mathematical property similar to the sum of delta functions.

\subsubsection*{Proof of statement on symmetry of poles}
Now we prove the statement that $-\omega_j$ and $\pm\omega_j^*$ are poles of $S(\omega)$ if $\omega_j$ is a pole. Without loss of generality, suppose $\omega_j$ is a $k$th-order pole in the first quadrant of the complex plane, then $S(\omega)$ can be expanded as a Laurent series at $\omega_j$ as
\begin{equation}
    S(\omega) = \frac{S_0}{(\omega-\omega_j)^k} + \cdots.
\end{equation}
Since $S(-\omega)=S(\omega)$, then one can obtain another Laurent series of $S(\omega)$ as
\begin{equation}
    S(\omega) = S(-\omega) = \frac{(-1)^kS_0}{(\omega+\omega_j)^k} + \cdots,
\end{equation}
which indicates that $-\omega_j$ is also a $k$th-order pole of $S(\omega)$. On the other hand, noticing that $S(\omega)$ takes real values on the real axis, i.e. $S(\omega)\in\mathbb{R},\,\forall\omega\in\mathbb{R}$, the Schwarz reflection principle states that $S(\omega^*)=[S(\omega)]^*$. This property implies another Laurent series of $S(\omega)$ as
\begin{equation}
    S(\omega) = [S(\omega^*)]^* = \frac{S_0^*}{(\omega-\omega_j^*)^k} + \cdots,
\end{equation}
which indicates that $\omega_j^*$ is also a $k$th-order pole of $S(\omega)$. Now that both $-\omega_j$ and $\omega_j^*$ are poles of $S(\omega)$, one can use the same method to prove that $-\omega_j^*$ is also a pole of $S(\omega)$.

\section{Coriolis Force}\label{appendix b}

In this appendix, we show that the Coriolis force on the experiment apparatus can only cause a pure imaginary valued cross-PSD $S_{a_xa_y}(\omega)$. In this case, the motion of the experiment apparatus in the $X-Y$ plane can be generally modelled as a Foucault pendulum under dissipation and random forces, then the equations of motion can be written as 2-dimensional Langevin equations
\begin{equation} \label{EOM for Coriolis}
\left\{\begin{aligned}
    \ddot{X} &= -\Omega_0^2X + 2\Omega_r\dot{Y} - \gamma_X\dot{X} + A_X(t), \\
    \ddot{Y} &= -\Omega_0^2Y - 2\Omega_r\dot{X} - \gamma_Y\dot{Y} + A_Y(t),
\end{aligned}\right.
\end{equation}
and $\Omega_r$ is a general Coriolis force term. The most commonly considered Coriolis force is the one induced by the Earth's rotation, which satisfies $\Omega_r=\Omega_{\rm earth}\sin\lambda\sim2\pi\sin\lambda/86400$\,Hz with $\lambda$ as the latitude of the experiment location. In frequency space, the dynamical equations become  
\begin{equation}
\left\{\begin{aligned}
    -\omega^2X(\omega) = -\Omega_0^2X(\omega) + 2i\omega\Omega_rY(\omega) - i\omega\gamma X(\omega) + A_X(\omega), \\
    -\omega^2Y(\omega) = -\Omega_0^2Y(\omega) - 2i\omega\Omega_rX(\omega) - i\omega\gamma Y(\omega) + A_Y(\omega).
\end{aligned}\right.
\end{equation}
The solution also has a form as \eqref{mechanical susceptibility}, with the mechanical susceptibility $\chi(\omega)=T/\text{det}T$ is given by
\begin{equation}
\begin{aligned}
    T &= \begin{pmatrix}
        \Omega_0^2-\omega^2+i\omega\gamma & 2i\omega\Omega_r \\
        -2i\omega\Omega_r & \Omega_0^2-\omega^2+i\omega\gamma
    \end{pmatrix}, \\
    \text{det}T &= \left(\Omega_0^2-\omega^2+i\omega\gamma\right)^2-4\omega^2\Omega_r^2.
\end{aligned}
\end{equation}
As a result, the PSDs $S_{XX}(\omega)$, $S_{YY}(\omega)$ and the cross-PSD $S_{XY}(\omega)$ can be written as the matrix form
\begin{widetext}
\begin{equation}
\begin{aligned}
    \begin{pmatrix}
        S_{XX}(\omega) & S_{XY}(\omega) \\
        S_{YX}(\omega) & S_{YY}(\omega)
    \end{pmatrix} &= S_0\chi(\omega)\chi^\dagger(\omega) 
        = \frac{S_0}{|\text{det}T|^2} \begin{pmatrix}
        (\Omega_0^2-\omega^2)^2+\omega^2(\gamma^2+4\Omega_r^2) & 4i\omega\Omega_r(\Omega_0^2-\omega^2) \\
        -4i\omega\Omega_r(\Omega_0^2-\omega^2) & (\Omega_0^2-\omega^2)^2+\omega^2(\gamma^2+4\Omega_r^2)
    \end{pmatrix}, \\
    |\text{det}T|^2 &= \left((\Omega_0^2-\omega^2)^2-\omega^2(\gamma^2+4\Omega_r^2)\right)^2+4\omega^2\gamma^2(\Omega_0^2-\omega^2)^2.
\end{aligned}
\end{equation}
\end{widetext}
Consequently, the cross-PSD $S_{a_xa_y}(\omega)=\omega^4S_{XY}(\omega)$ is 
\begin{equation}
    S_{a_xa_y}(\omega) = \omega^4S_0\frac{4i\omega\Omega_r(\Omega_0^2-\omega^2)}{|\text{det}T|^2}.
\end{equation}
It is notable that the cross PSD $S_{a_xa_y}(\omega)$ is a pure imaginary valued function, which indicates that the cross-correlation between $a_x$ and $a_y$ is determined by the out-of-phase components. However, as is pointed out in \eqref{sigma simplified}, the variance of the phase fluctuation only relies on the in-phase components of $S_{a_xa_y}(\omega)$, so the cross term of $a_x$ and $a_y$ determined by the dynamical equation \eqref{EOM of apparatus} exactly has no contribution to the phase fluctuation.

\section{Simulation of Inertial Noise}\label{appendix d}

\begin{figure*}
    \centering
    \begin{minipage}{0.45\textwidth}
        \includegraphics[scale=0.4]{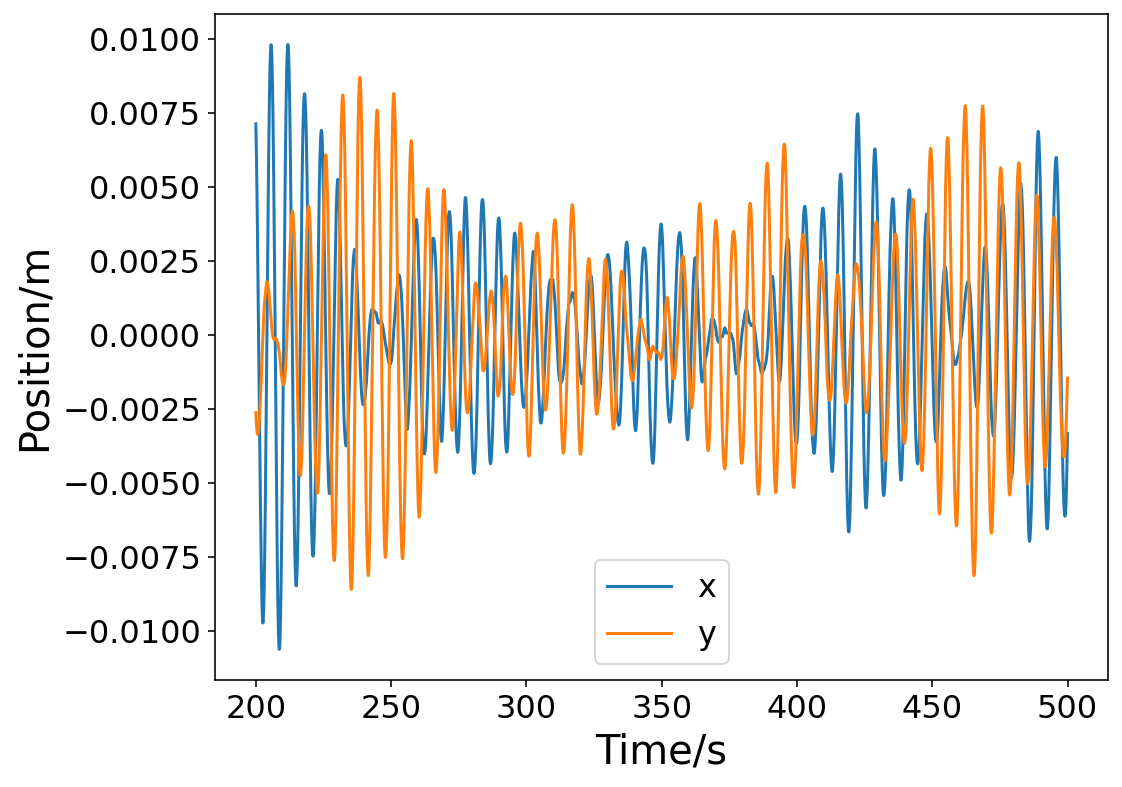}
    \end{minipage}
    \hfill
    \begin{minipage}{0.45\textwidth}
        \includegraphics[scale=0.4]{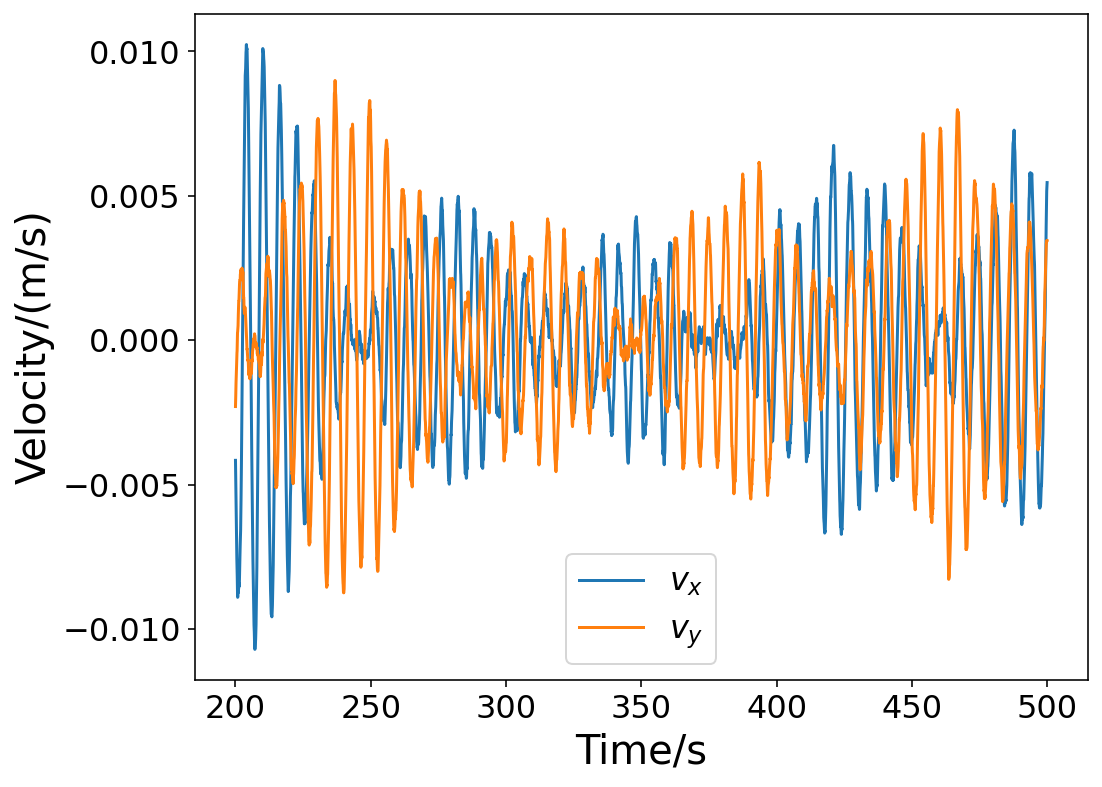}
    \end{minipage}
    \caption{\small The motion of the experiment apparatus. The parameters are chosen as $\Omega_0=1$\,Hz, $\gamma=0.01$\,Hz, $k=0.3\,\text{Hz}^2$, and the iteration step is chosen as 0.05\,s. Since the initial conditions are chosen as $X(0)=Y(0)=v_x(0)=v_y(0)=0$, the beginning of the simulation is near zero and unstable, so we cut off the beginning data for data process.}
    \label{xyt}
\end{figure*}

In this appendix, we show the details of the simulation of the inertial noise and the data process of the simulation result. We use the 4th order Runge-Kutta method to simulate the dynamical equation \eqref{EOM of apparatus}. To start with, we rewrite the second order differential equations as first order differential equations by introducing $v_{X, Y}\equiv\dot{X}, \dot{Y}$, then the dynamical equation \eqref{EOM of apparatus} becomes 
\begin{equation}
\left\{\begin{aligned}
    \dot{X} &= v_X, \\
    \dot{Y} &= v_Y, \\
    \dot{v}_X &= -\Omega_0^2X + 2\Omega_rv_Y - \gamma_X\dot{X} + A_X(t), \\
    \dot{v}_Y &= -\Omega_0^2Y - 2\Omega_rv_X - \gamma_Y\dot{Y} + A_Y(t).
\end{aligned}\right.
\end{equation}
These equations can be mathematically written as $\dot{u}_i(t) = f_i[u_j(t), A_m(t)]$ with $i, j=X, Y, v_{X, Y}$ and $m=X, Y$. The initial conditions are chosen $u_i(0)=0$ because we want to focus on the driven effects of the noise $A_{X, Y}(t)$. Then the Runge-Kutta method intimates the iterative equations
\begin{equation}\label{iteration eq}
    u_i(t_n+\Delta t) = u_i(t_n) + \frac{\Delta t}{6}\left(k_{i1}+2k_{i2}+2k_{i3}+k_{i4}\right),
\end{equation}
with the slope parameters $k_{i1}\sim k_{i4}$ as
\begin{equation}\label{slope para}
\begin{aligned}
    k_{i1} &= f_i[u_j(t_n), A_m(t_n)], \\
    k_{i2} &= f_i[u_j(t_n)+k_{i1}\Delta t/2, A'_m], \\
    k_{i3} &= f_i[u_j(t_n)+k_{i2}\Delta t/2, A'_m], \\
    k_{i4} &= f_i[u_j(t_n)+k_{i3}\Delta t, A_m(t_{n+1})]. 
\end{aligned}
\end{equation}
Remarkably, we choose $A'_m$ as $(A_m(t_n)+A_m(t_{n+1}))/2$ because the noise is a stochastic process rather than a continuous function, otherwise, it is usually chosen as $A'_m=A_m(t_n+\Delta t/2)$ according to the standard Runge-Kutta method. In fact, for the Ornstein-Uhlenbeck process, more precise simulation algorithms are available~\cite{PhysRevE.54.2084, Cai_2012}. However, the simulation algorithm in this paper is sufficiently accurate for our purposes.

Based on the Runge-Kutta iteration equations \eqref{iteration eq} and \eqref{slope para}, we simulated the evolution of the experiment apparatus, shown as Fig.\,\ref{xyt}. Note that the behaviours of $X(t)$, $Y(t)$ are similar to their velocities $v_{X, Y}(t)$ with a $\pi/2$-phase difference. 

Based on the simulation result shown in Fig.\,\ref{xyt}, we compute the PSD and cross-PSD by the fast Fourier transform (FFT) by the SciPy package. The simulated noise is divided into 10 segments with an overlap length of half the segment length between adjacent segments. The simulated noise also multiplies a Hann window to avoid the frequency leakage problem. Then the PSD and cross-PSD shown in Fig.\,\ref{PSD} can be computed by the FFT.

% \onecolumngrid
% \bibliographystyle{apsrev}
% \bibliography{literature} 

\end{document}